\documentclass[aip,reprint]{revtex4-1}

\usepackage{grffile}

\usepackage{amsmath,amsfonts,amscd}
\usepackage{amsmath}
\usepackage{dcolumn}
\usepackage{graphicx}
\usepackage{color}
\usepackage{color}
\usepackage{transparent}
\usepackage[utf8]{inputenc}
\usepackage[T1]{fontenc}
\usepackage{mathptmx}

\newcommand{\R}{\mathbb{R}}
\newcommand{\C}{\mathbb{C}}

\newcommand{\Z}{\mathbb{Z}}

        \newcommand{\mc}[1]{\mathcal{#1}}
        \newcommand{\mb}[1]{\mathbb{#1}}

        \newcommand{\lp}{\left}
        \newcommand{\rp}{\right}
                \newcommand{\la}{\lp\langle}
        \newcommand{\ra}{\rp\rangle}
        \newcommand{\beq}{\begin{equation}}
        \newcommand{\eeq}{\end{equation}}
        \newcommand{\ba}{\begin{align}}
        \newcommand{\ea}{\end{align}}
        
        \newcommand{\p}{\partial}
        \newcommand{\ri}{\mathrm{i}}
        
        \newcommand{\rlin}{\mathrm{lin}}
        
        \newcommand{\re}{\mathrm{e}}
        
         \newcommand{\rf}{\mathrm{f}}

        \renewcommand{\theenumi}{(\roman{enumi})}

\draft 

\begin{document}


\title{Transverse modulational dynamics of quenched patterns} 



\author{Sierra Dunn}
\affiliation{Department of Mathematics and Statistics, Mount Holyoke College, 415A Clapp Laboratory, South Hadley, MA, 01075 USA;}
\author{Ryan Goh}
\email[]{rgoh@bu.edu} 
\affiliation{Department of Mathematics and Statistics, Boston University, 665 Commonwealth Ave., Boston,  MA 02215, USA;}
\author{Benjamin Krewson}
\affiliation{Department of Mathematics and Statistics, Boston University, 665 Commonwealth Ave., Boston,  MA 02215, USA;}



\date{\today}

\begin{abstract}
We study the modulational dynamics of striped patterns formed in the wake of a planar directional quench. Such quenches, which move across a medium and nucleate pattern-forming instabilities in their wake, have been shown in numerous applications to control and select the wavenumber and orientation of striped phases. In the context of the prototypical complex Ginzburg-Landau and Swift-Hohenberg equations, we use a multiple-scale analysis to derive a one-dimensional viscous Burgers' equation which describes the long-wavelength modulational and defect dynamics in the direction transverse to the quenching motion, that is along the quenching line. We show that the wavenumber selecting properties of the quench determines the nonlinear flux parameter in the Burgers' modulation equation, while the viscosity parameter of the Burgers' equation is naturally determined by the transverse diffusivity of the pure stripe state. We use this approximation to accurately characterize the transverse dynamics of several types of defects formed in the wake, including grain boundaries and phase-slips.
\end{abstract}


\maketitle 

\begin{quotation}
Directional quenching is a novel way to harness self-organized pattern-forming processes in a variety of systems. Here an external mechanism rigidly progresses across a medium, exciting pattern-forming instabilities in its wake. While recent work has sought to understand how the orientation and wavenumber of striped patterns are selected by the quench, little has been done to understand the dynamics of defects and modulations of such patterns. Focusing on the interfacial dynamics of the patterned front just behind the quench, we derive a viscous Burgers' modulation equation to understand slowly-varying modulational dynamics in the direction perpendicular, or transverse, to the quenching motion. Crucially, we find that the selected wavenumber in the direction of quenching motion determines the parameters of the viscous Burgers' equation. We evidence the ubiquity of this modulational approximation by characterizing several types of wavenumber defects in quenched versions of the prototypical Complex Ginzburg-Landau and Swift-Hohenberg equations.

\end{quotation}

\section{Introduction}\label{s:intro}

Directional quenching has arisen as a novel way to harness, mediate, and control pattern forming instabilities in diverse application areas. Generally, some sort of external mechanism, possibly controlled by the experimenter, travels across the domain initiating a pattern-forming instability in its wake. One then hopes to control the shape, size, and orientation of the pattern by altering the speed and structure of the quench. Some examples include ramped fluid flows (Ref. \onlinecite{riecke1986pattern}), solidification problems in crystal formation (Ref. \onlinecite{zigzageutectic}) , and light-sensing reaction-diffusion experiments (Refs. \onlinecite{PMID:30860212,CDIMA}). 
Additionally, directionally quenched systems serve as a prototype and testbed to understand how spatio-temporal heterogeneities and growth processes affect pattern-forming systems.  Most works have studied the existence of stripe-forming front solutions in the wake of a quench, in particular focusing on how the quenching speed and shape affect the orientation and bulk wavenumber of the far-field pattern.  Furthermore, a few works have studied (in)stability of these front solutions in specific systems (Ref. \onlinecite{MR3958766,goh2020spectral}). See (Ref. \onlinecite{goh21a,goh2023growing}) for recent reviews of these works as well as references to other application areas. 

\begin{figure*}
\centering
(a)\hspace{-0.04in} \includegraphics[trim = 1.1cm 0.0cm 0.5cm 0.0cm,clip,width=0.3\textwidth]{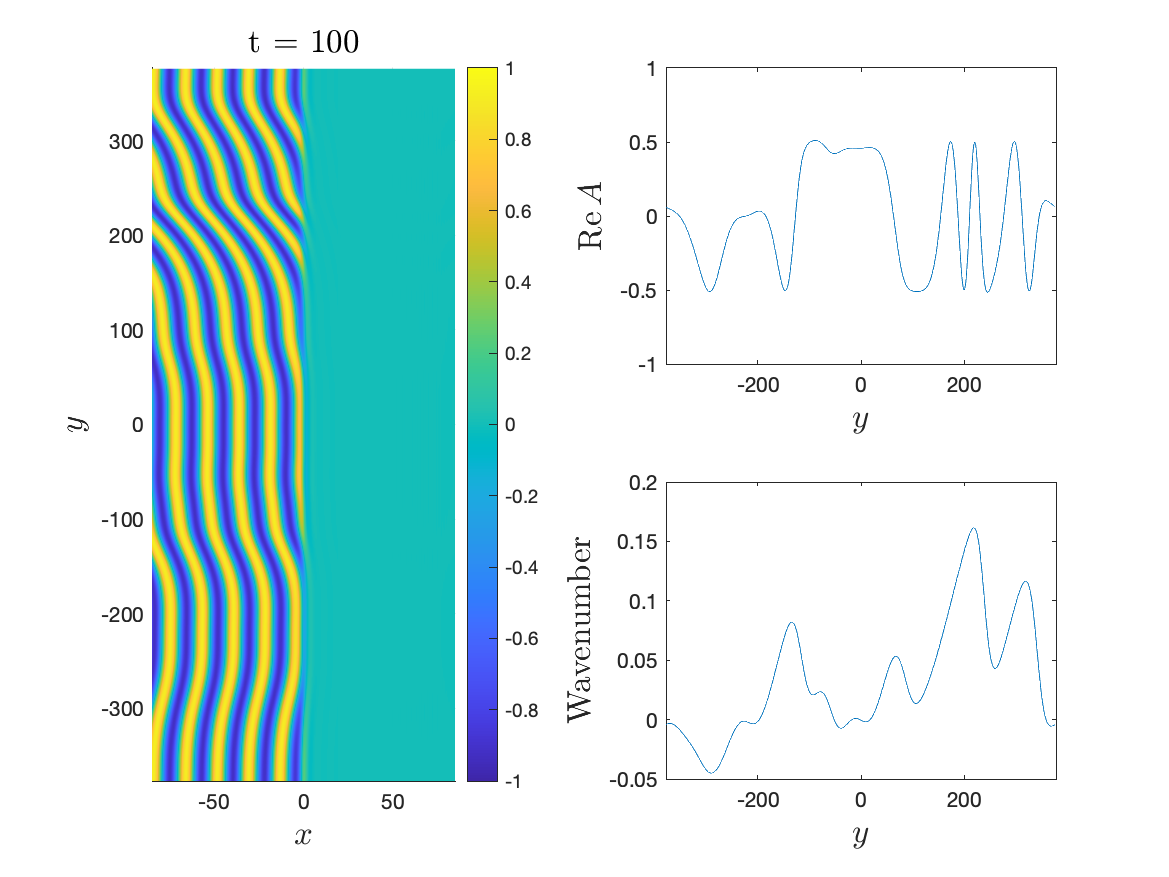}\hspace{-0.2in}
(b)\hspace{-0.04in} \includegraphics[trim = 1.1cm 0.0cm 0.5cm 0.0cm,clip,width=0.3\textwidth]{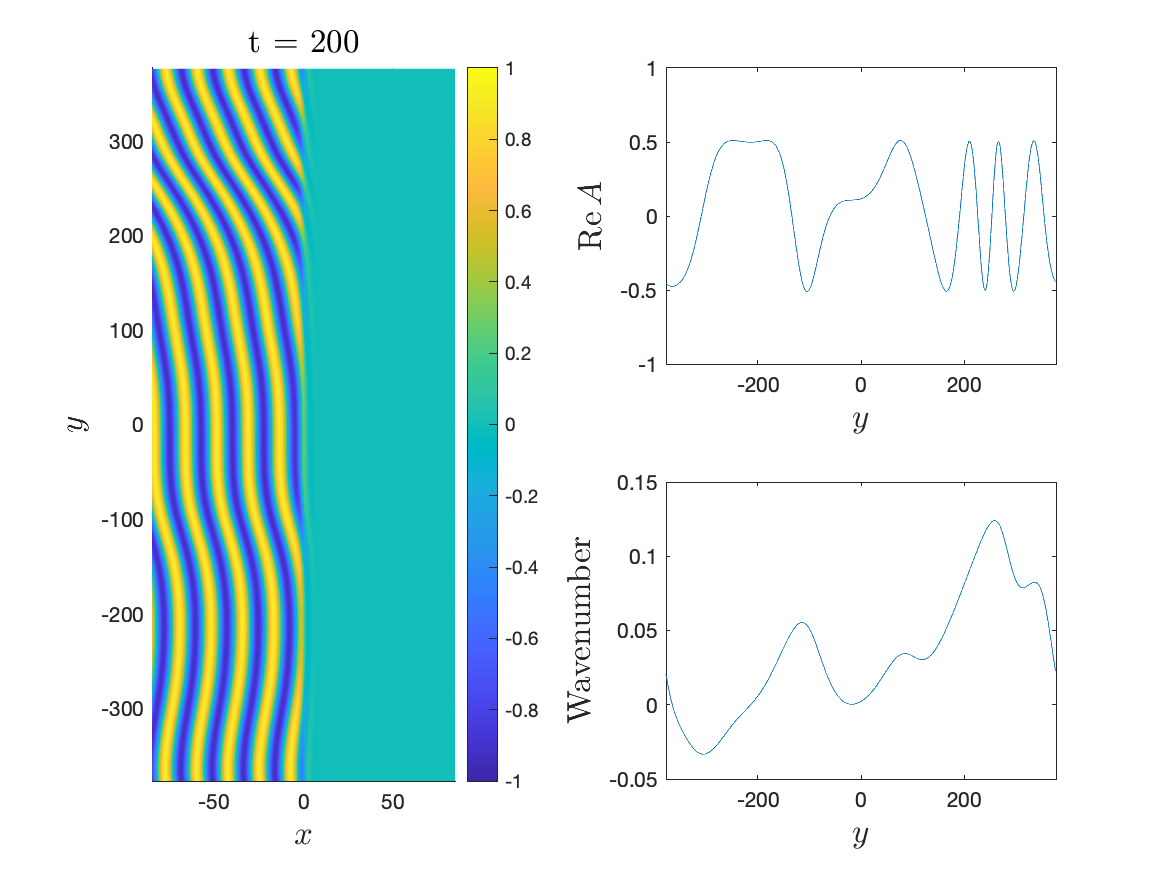}\hspace{-0.2in}
(c)\hspace{-0.04in} \includegraphics[trim = 1.1cm 0.0cm 0.5cm 0.0cm,clip,width=0.3\textwidth]{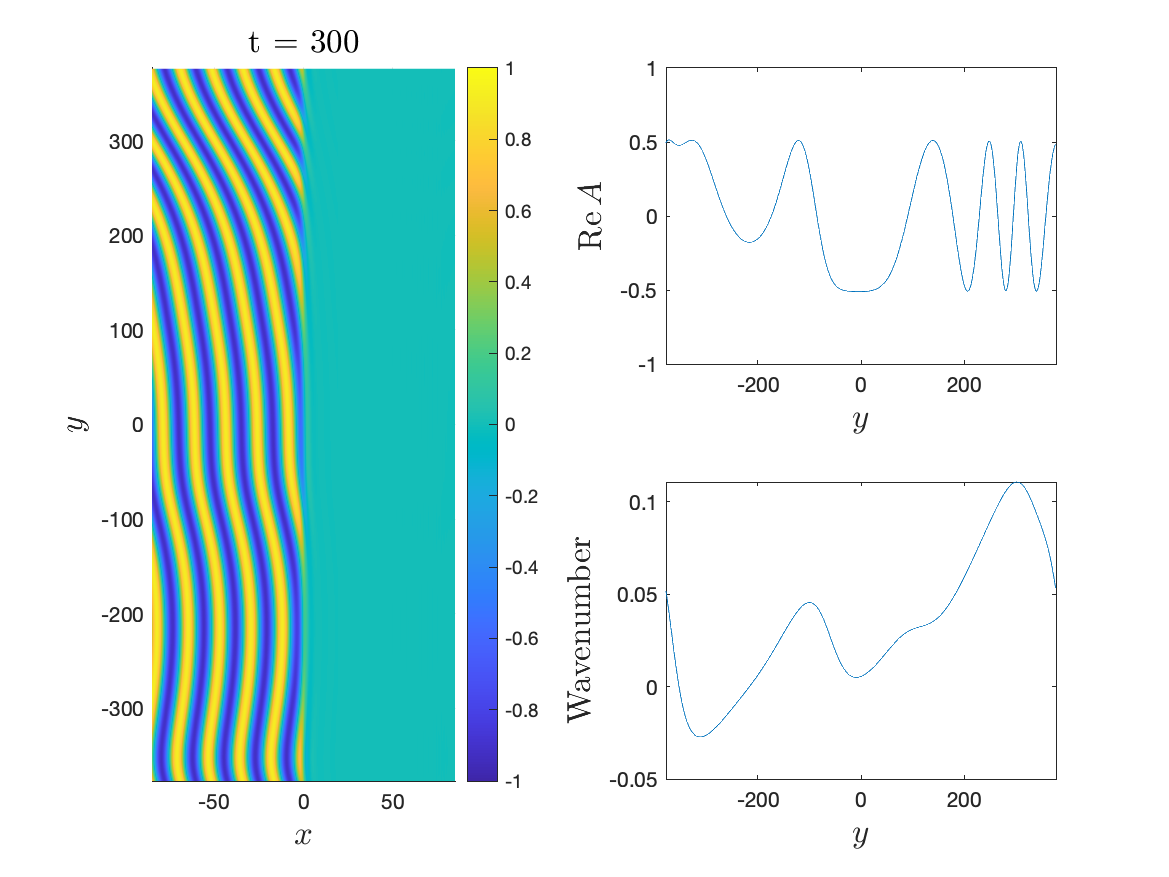}\hspace{-0.2in}
\caption{Evolution of 2D quenched pattern in \eqref{e:cgl-c}, with $\alpha = 3, \gamma = 1, c = 2.5$, from small random initial data at times $t = 100,200,300$ (a) - (c) respectively; left subplots depict $\mathrm{Re}\,A(\xi,y,t)$, upper right plots $\mathrm{Re}\,A(\xi_0,y,t)$ for $\xi_0 = -1$, while lower right plots the local transverse wavenumber, calculated as $\mathrm{Im}\, \frac{A_y}{A}$ for $\xi = \xi_0$ fixed.  }\label{f:rand}
\end{figure*}

In comparison, relatively little is known about the dynamics, defects, and interactions of striped patterns formed in different sub-domains behind the quench. This becomes a question of interest, for example, when one studies the evolution of a quenched system starting with small fluctuations of the homogeneous background state. Here, patches of large amplitude patterns interact through defects which move along the quenching line. See for example Figure \ref{f:rand} which depicts the evolution of the quenched domain from random perturbation of the trivial state in the complex Ginzburg-Landau equation.

This work considers quenched patterns in two spatial dimensions where the quench rigidly propagates in the horizontal direction. Previous works have shown that such quenches select the horizontal wavenumber $k_x$, and thus the temporal frequency $\omega$, of the asymptotic pattern. In particular, they are determined by the quenching speed $c$ and the transverse wavenumber $k_y$.   We use a formal multiple-scales analysis to derive a reduced one-dimensional model for transverse modulations of striped patterns, that is we consider vertical modulations along the quenching line. 
As they determine the far-field pattern, we focus on solution dynamics just behind the quenching line. We show that a one-dimensional viscous Burgers' equation accurately predicts the dynamics of slowly-varying, small amplitude wavenumber modulations. 
Most strikingly, we find that the selection of a unique horizontal wavenumber $k_x$ for a given speed $c$  determines the viscosity and nonlinear flux parameters in the associated viscous Burgers' equation.   We first demonstrate our approach  through asymmetric grain boundary and phase-slip examples in the prototypical complex Ginzburg Landau equation. We then show its applicability in the Swift-Hohenberg equation, studying similar types of defects. We expect such modulation equations will predict transverse dynamics in many other quenched systems where the asymptotic pattern is diffusively stable (Ref. \onlinecite{schneider} ). Finally, while we mostly focus on transverse modulations of vertically independent stripes, with $k_y = 0$ so they are oriented parallel to the quench, we expect our results to apply to slowly-varying modulations of obliquely oriented stripes as well.


\section{Prototypical Example: the quenched Complex Ginzburg-Landau equation}

\subsection{Quenched stripes}
\begin{figure}
\centering
\hspace{-0.0in}(a)\hspace{-0.04in}  \includegraphics[trim = 0.0cm 0.0cm 0.0cm 0.0cm,clip,width=0.35\textwidth]{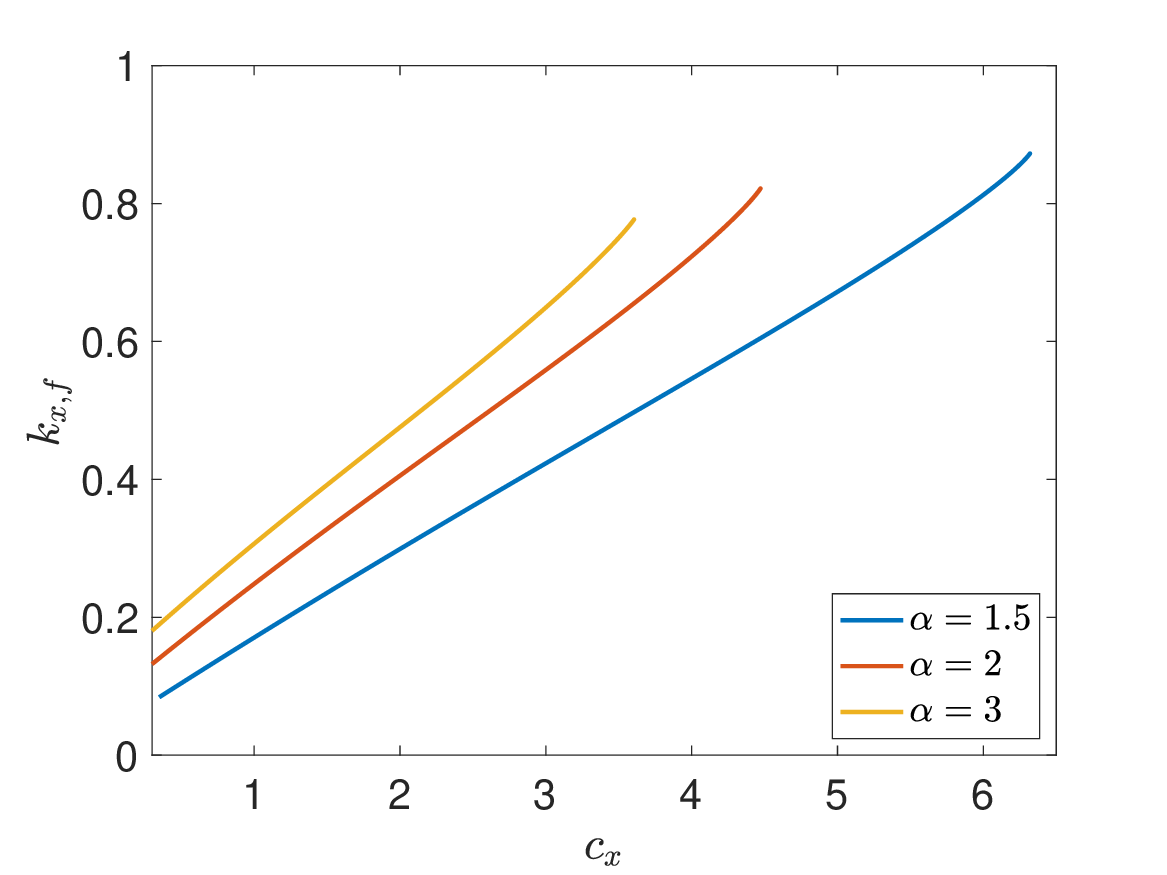}\hspace{-0.0in}\\
\hspace{-0.2in}(b)\hspace{-0.04in} \includegraphics[trim = 0.0cm 00cm 00cm 0.0cm,clip,width=0.32\textwidth]{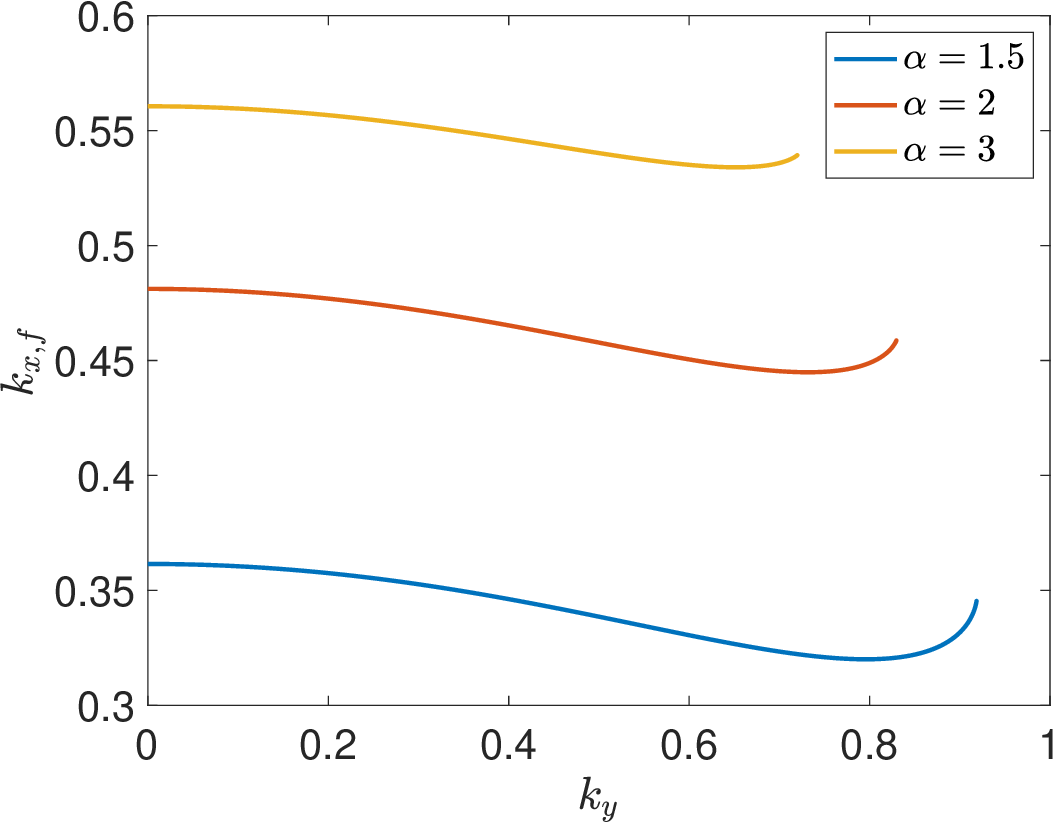}\hspace{-0.0in}
\caption{Wavenumber selection curves with $\gamma = 1$ and $\alpha =1.5,2,3$. Plot of the wavenumber selection curves $k_{x,\mathrm{f}}$ against $c$ for $k_y = 0$ fixed (a), and against $k_y$ for $c = 2.5$ fixed (b). Note the quadratic behavior near $k_y = 0$.}\label{f:auto}
\end{figure}

To introduce our approach, we consider the complex Ginzburg-Landau (CGL) equation (Ref. \onlinecite{CGL_Lambda_Stability}) with cubic supercritical nonlinearity, posed in the plane,
\begin{eqnarray}\label{e:cgl}
A_t &= (1+\ri \alpha) \Delta A + \chi(x - ct)A - (1+\ri\gamma)A|A|^2 \\
&\quad A\in \C,\quad (x,y)\in \R^2, \quad \alpha,\gamma\in \R,\notag
\end{eqnarray}
with directional quenching heterogeneity, $\chi(\xi) = -\mathrm{sign}(\xi)$, a step-function which rigidly propagates with speed $c\geq0$ and which renders the trivial state $A = 0$, which is stable for $x-ct>0$, into an unstable state for $x-ct <0$. We transform into a co-moving frame of speed $c$ in the horizontal direction, setting $\xi = x - ct$, to obtain
\begin{align}\label{e:cgl-c}
A_t &= (1+\ri \alpha)(\partial_\xi^2 + \partial_y^2) A + c A_\xi + \chi(\xi)A - (1+\ri\gamma)A|A|^2. 
\end{align}
 Due to the invariance of the equation under the gauge action $A\mapsto \re^{i\theta} A$, the homogeneous version of \eqref{e:cgl-c} with $\chi\equiv1$ has explicit spatially periodic relative equilbria $A_\mathrm{p}(\xi,y,t;k_x,k_y) = r\re^{i(\omega t + k_x \xi + k_y y)}$ with respect to this symmetry action. Furthermore, the amplitude $r$ and wavenumber $k^2 = k_x^2 + k_y^2$ satisfies the following nonlinear dispersion relation in the co-moving frame
\beq\label{e:cgl-disp}
r^2 = 1-k^2,\quad \omega = (\gamma - \alpha)k^2 + ck_x - \gamma.
\eeq
Once again due to the gauge invariance, the simplest stripe forming front solutions of \eqref{e:cgl-c} take the form
\beq
A(\xi,y,t) = \re^{\ri(k_y y + \omega t)}A_\mathrm{f}(\xi;c,k_y), 
\eeq
where $A_\mathrm{f}$ is a function of the co-moving frame variable $\xi$ and the system parameters $(c,k_y)$, and solves the following traveling wave ODE with corresponding asymptotic boundary conditions
\begin{align}\label{e:tw}
0 &= (1+\ri\alpha)(\partial_\xi^2 - k_y^2)A_\mathrm{f} + cA_{\rf,\xi} + (\chi(\xi)-\ri \omega)A_\rf\notag\\
&\qquad\qquad\qquad\qquad\qquad\qquad\qquad - (1+\ri\gamma)A_\rf|A_\rf|^2,\\
0 &= \lim_{\xi\rightarrow -\infty} \left| A_\mathrm{f}(\xi) - r \re^{\ri k_x \xi}\right|,\qquad
0 = \lim_{\xi\rightarrow + \infty} A_\mathrm{f}(\xi).
\end{align}
\noindent To summarize, $A_\mathrm{f}$ connects the stable trivial state ahead of the quench to a periodic pattern with horizontal wavenumber $k_x$. This in turn gives a solution of the full PDE which connects a striped pattern with wavevector $(k_x,k_y)$ to the trivial state ahead of the quench. The work (Ref. \onlinecite{gs1}) rigorously established the existence of such fronts for $k_y = 0$ in the fast growth regime where $c\lesssim c_\rlin:=2\sqrt{1+\alpha^2}$, the linear spreading speed of fronts invading into the homogeneous unstable state for $\chi\equiv1$ (Ref. \onlinecite{vS}). It  showed that the temporal frequency $\omega$, and thus the horizontal wavenumber $k_x$ is determined, or ``selected," by the quenching speed $c$, giving leading order expansions for this dependence. Further, since the term involving $k_y$ is a regular perturbation, we expect a family of front solutions, smoothly dependent on $k_y^2$, to persist for $k_y\sim0$. Hence the frequency $\omega$ and horizontal wavenumber $k_x$ will also be selected by $k_y^2$. In sum, these quantities can be written locally as graphs $\omega_\mathrm{f}(c,k_y), k_{x,\mathrm{f}}(c,k_y)$ over $(c,k_y)$-space. We denote the corresponding family of front solutions as $A_\mathrm{f}(\xi;k_y,c)$. Figure \ref{f:auto} gives numerical continuation results using AUTO07p (Ref. \onlinecite{doedel2007auto}) which continue fronts $A_\rf$ and horizontal wavenumber $k_x$ in both $c$ and $k_y$ for several values of $\alpha$. We refer to the corresponding surface in $(c,k_y,k_x)$-space  as the \emph{moduli space of quenched stripes} (Ref. \onlinecite[\S 5.3]{goh2023growing}) as it organizes which type(s) of stripes can be selected for a given quenching speed.  The rigorous existence of such fronts for other values of $c$ and $k_y\neq 0$ is the subject of current work. 

For the remainder of the work we shall fix $c>0$ such that a traveling-front solution of \eqref{e:tw} exists for all $k_y$ close to 0, and hence we suppress the dependence of $c$ in our notation. We consider parameters $\alpha,\gamma$ such that stripes are Benjamin-Feir stable, $1+\alpha\gamma>0$. Using the dependence of $k_x$ on $k_y$, the nonlinear dispersion relation then takes the form
\beq\label{e:cgl_disp}
\omega_\rf(k_y) = (\gamma - \alpha)(k_{x,\mathrm{f}}(k_y)^2 + k_y^2) + c k_{x,\mathrm{f}}(k_y) - \gamma.
\eeq
Since $k_{x,\rf}$ is smoothly dependent on $k_y^2$, a simple calculation gives that both $\frac{\partial}{dk_y} k_{x,\rf}(0) = \frac{\partial}{dk_y} \omega_\rf(0) =  0$, yielding the expansion
$$
k_{x,\mathrm{f}}(k_y;c) = k_{x,\mathrm{f}}(0) + \beta_2 k_y^2 + \mathcal{O}(k_y^4),\qquad k_y\sim0,
$$
for some constant $\beta_2$. 
See Figure \ref{f:auto} for a depiction of this quadratic dependence near $k_y = 0$. As it will be critical in our modulation equation, we also calculate
\beq\label{e:wnl}
\partial_{k_y}^2 \omega_\rf(0) = 2(\gamma - \alpha) + 2\p_{k_y}^2k_{x,\mathrm{f}}(0)\left( 2(\gamma - \alpha) k_{x,\mathrm{f}}(0) - c\right).
\eeq

Initiating \eqref{e:cgl-c} from small random initial data with speed $c = 2.5$, one observes the formation of patches of coherent stripes, oriented with weak oblique angle to the quench interface.  In between these patches lie a variety of defects, including grain boundaries and dislocations; see Figure \ref{f:rand}. Taking a cross-section in $y$ for some $\xi_0<0$ just behind the quench location at $\xi = 0$, one observes wavenumber dynamics similar to 1-D systems. 

\subsection{Transverse modulations}
To study the dynamics of small amplitude, long wavenlegnth transverse modulations in the $y$-direction, we adapt the approach of (Ref. \onlinecite{dmwt}) which considers stripes in 1-D; see also (Ref. \onlinecite[\S II.G]{CGL_Lambda_Stability}). 
We look for slow transverse modulations of front solutions of \eqref{e:cgl-c} by introducing a slowly-varying transverse phase modulation function $\Phi(Y,T)$ dependent on the slow variables $Y = \delta y, \, T = \delta^2 t$ for some small parameter $0<\delta\ll1$, and form the ansatz
$$
A(\xi,y,t) = \re^{\ri(\Phi(Y,T) - \omega_\rf t)}A_\mathrm{f}(\xi;\delta \Phi_Y(Y,T)),
$$
where $\Psi:= \p_Y\Phi$ gives the slowly-varying transverse wavenumber modulation. Inserting this ansatz into \eqref{e:cgl-c} and its associated complex-conjugate equation, we then collect terms of the same order in $\delta$, and fix  $\xi = \xi_0$, a location just behind the quench with $-1\leq \xi_0 <0.$ At $\mathcal{O}(1)$ in $\delta$ we obtain the traveling wave equation \eqref{e:tw} for the front, at $\mathcal{O}(\delta)$ we obtain an equation for the kernel of the associated linearized equation, satisfied by the derivative of the front solution along the gauge action, and finally at $\mathcal{O}(\delta^2)$  we obtain the following viscous Burgers equation for the transverse wavenumber,
\begin{equation}\label{e:vb}
\Psi_T = \frac{\lambda_\mathrm{lin}''(0)}{2}\Psi_{YY} + \frac{\omega_\rf''(0)}{2}(\Psi^2)_Y,
\end{equation}
where $\omega_\rf''(0)$ is given in \eqref{e:wnl} above and $\lambda_\mathrm{lin}''(0)\approx 2(1+\alpha\gamma)$ 
gives the effective diffusivity of transverse perturbations of the parallel striped state,  obtained by perturbing the pure striped solution in the $y$-direction with the ansatz,
\begin{equation}
A(\xi,y,t) = A_p(\xi,y,t;k_x,k_y) + \re^{\ri(k_y y + \omega t)} \left[a_1 \re^{\lambda t + \nu y} + a_2 \re^{\lambda t - \nu y} \right],\label{e:cgl_modan}
\end{equation}
collecting leading order terms in $a_1,a_2$, and solving to obtain
\beq
\lambda = \lambda_\mathrm{lin}(\nu) = -c_g \nu + \lambda_\mathrm{lin}''(0)\nu^2 + \mc{O}(\nu^3),\,\, \nu\in\ri\R,
\eeq
where 
$$
c_g = \omega_\rf'(0) = (\gamma - \alpha)\left(2k_{x,\mathrm{f}}(0) k_{x,\mathrm{f}}'(0)) + c k_{x,\mathrm{f}}'(0) \right),
$$ gives the transverse group velocity for parallel stripes with $k_y = 0$. 
Since $k_{x,\mathrm{f}}'(0) = 0$ one readily calculates that $c_g = \omega_\rf'(0) = 0$.  For more details of these calculations, see Appendix \ref{a:1}. 

We remark that the modulation ansatz \eqref{e:cgl_modan} will not be accurate in the far-field as we modulate the front uniformly in $\xi$. Despite this, since the interfacial dynamics will be convected into the bulk in the co-moving frame traveling with the quench, we expect our modulation to give good qualitative predictions of the far-field dynamics; see Sec. \ref{s:conc} for brief discussion on possible extensions of our work addressing this.  

\subsubsection{Numerical approach}
In the following examples, we give comparisons between the numerically measured transverse wavenumber dynamics of \eqref{e:cgl-c} and the corresponding numerical solutions of the viscous Burgers' equation \eqref{e:vb}. We simulate the quenched CGL equation using a Galerkin spectral discretization and the fast Fourier transform in both space directions on a periodic domain, $(x,y)\in[-L_x/2, L_x/2]\times [-L_y/2,L_y/2]$ with $L_x =30\pi, L_y = 120\pi$ and $N_x = 2^8, N_y = 2^{12}$ modes in the $x$ and $y$ direction respectively. The quench damping level was strengthened to $\chi = -3$  for $x>L_x/4$ to surpress fluctuations coming from the periodic boundary conditions and prepare a near homogeneous state close to $A = 0$ at the quenching line. The 4th order Runge-Kutta exponential time-differencing algorithm of (Ref. \onlinecite{kassam2005fourth}) was used to time step with step sizes ranging from $dt = 0.1$ to $dt = 0.0025$; for most figures $dt = 0.0025$ was used. Numerical solutions of the viscous Burgers equation \eqref{e:vb} were solved in the same manner, with spectral decomposition on the periodic computational domain $Y\in\delta[-L_y/2,L_y/2],\, N_y = 2^{12}$ and exponential time-differencing in $T$ with time-steps $\delta^2 dt$.  Computations were performed in MATLAB using both CPU and GPU computations.

\subsubsection{Source-sink transverse defect pair}
\begin{figure}
\centering
(a)\hspace{-0.04in}\includegraphics[trim = 1cm 0.0cm 0.0cm 0.0cm,clip,width=0.35\textwidth]{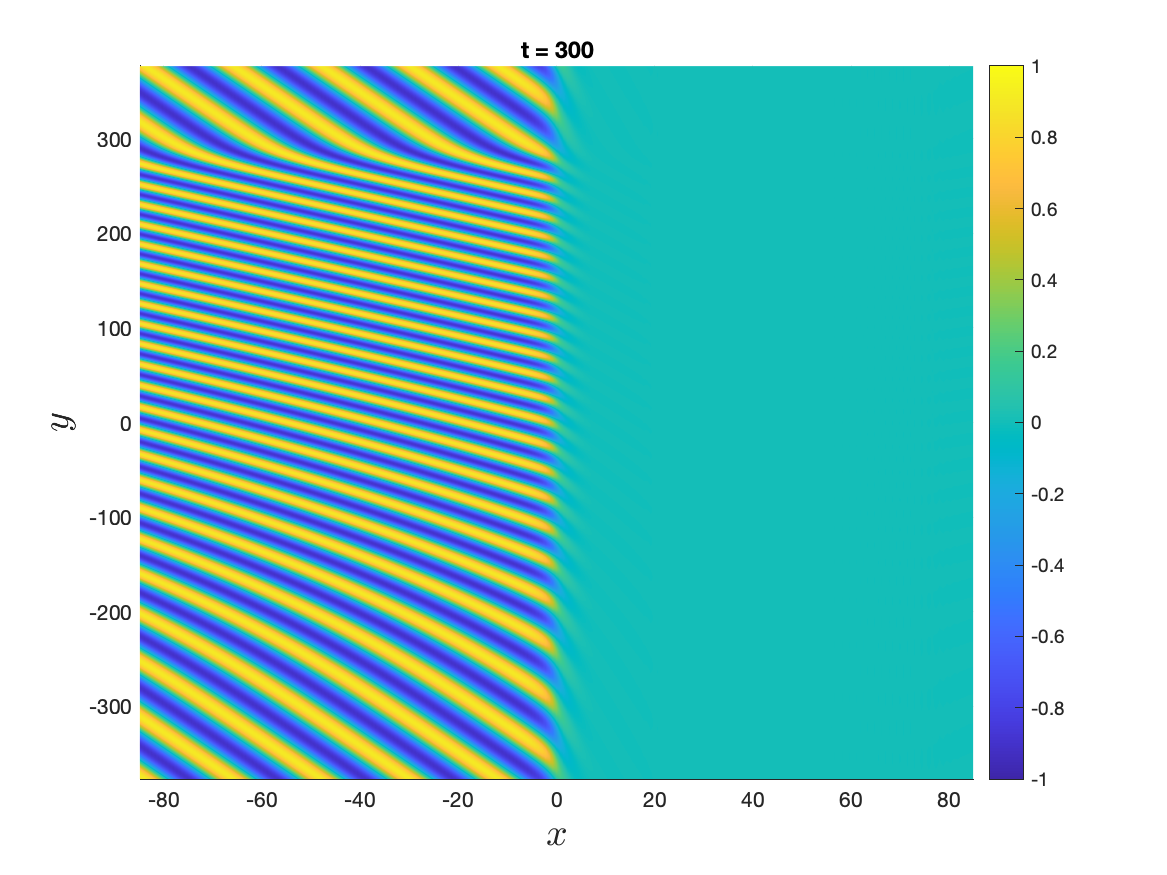}\hspace{-0.2in}\\
(b)\hspace{-0.04in} \includegraphics[trim = 0.0cm 0.0cm 0.0cm 0.0cm,clip,width=0.33\textwidth]{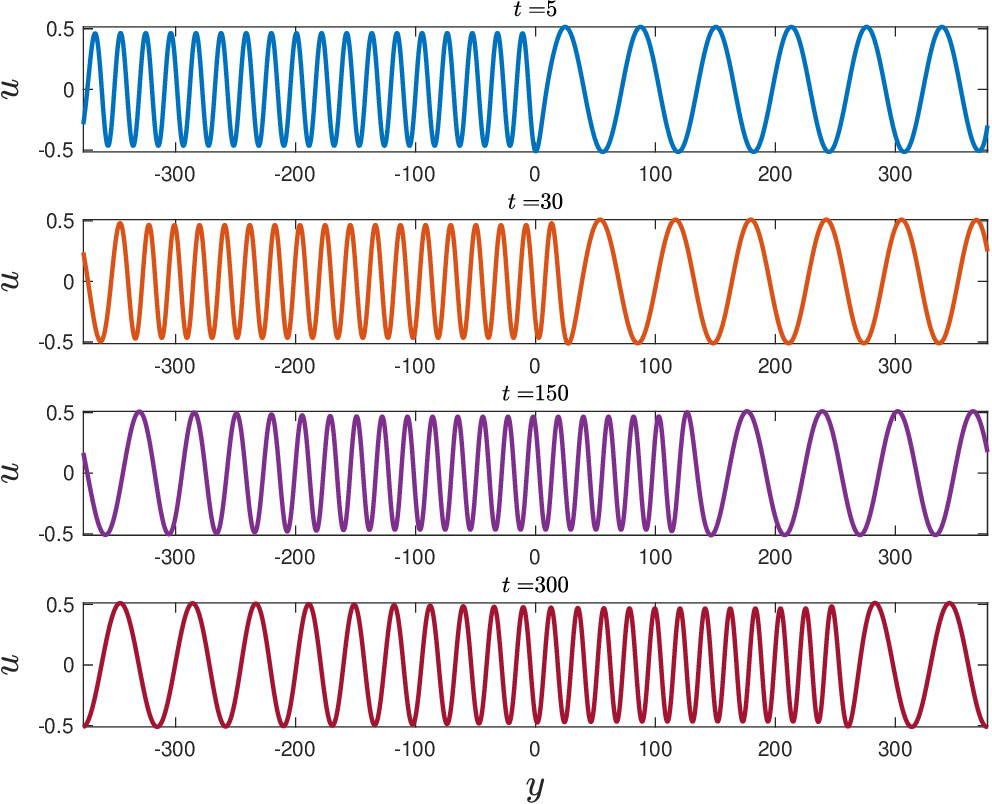}\hspace{-0.05in}\\
(c) \hspace{-0.04in}\includegraphics[trim = 0.2cm 0.0cm 0.0cm 0.0cm,clip,width=0.35\textwidth]{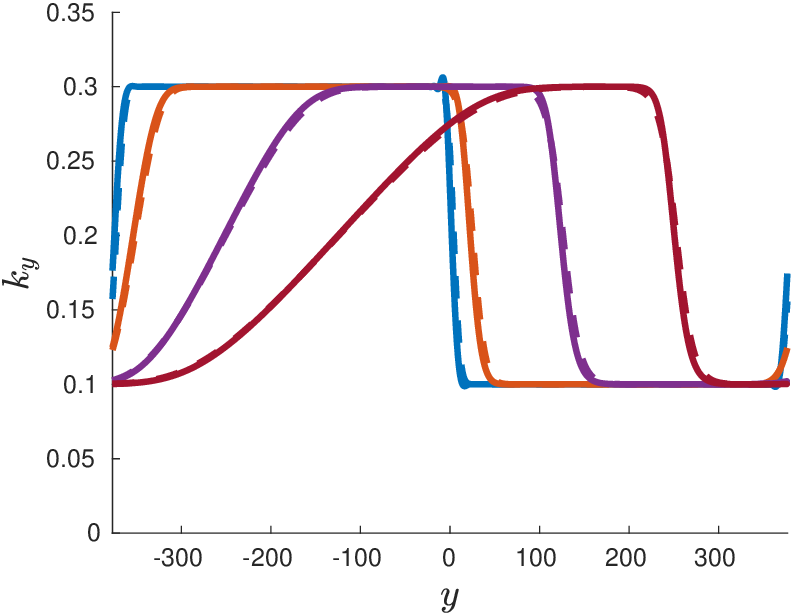}\hspace{-0.2in}
\caption{Source-sink defect pair with $\alpha = 3,\gamma = 1, c = 2.5$ obtained from initial condition \eqref{e:cglsrc-sk} which connects stripes solutions with transverse wavenumbers $k_{y,-} = 0.3$, $k_{y,+} = 0.1$ and $\delta = 0.1$;  (a): Plot of $\mathrm{Re}\, A(\xi,y,t = 300)$ of \eqref{e:cgl-c}. 
Note the source defect initially lies on the upper and lower domain boundaries. (b): plot of the cross-section of $\mathrm{Re}\, A$ along the line $\xi = \xi_0 = -1$ for a range of times $t = 5 \text{(blue)}, 30 \text{(orange)}, 150 \text{(purple)}, 300 \text{(dark red)}.$  (c): Comparison of the local transverse wavenumber $\mathrm{Im}\, \partial_y A/A$ along the line $\xi = \xi_0 $ (solid), with the appropriately rescaled numerical solution of the viscous Burgers' equation \eqref{e:vb}, $\delta \Psi(\delta y, \delta^2 t)$ (dashed) with $\delta = 0.1$, for same range of times. The nonlinear Burgers' parameter was calculated from data in Fig. \ref{f:auto}b and \eqref{e:wnl} as $\omega''_\rf(0) = 4.2126...$. }
 \label{f:sink-source}
\end{figure}

As a case study, we study the transverse wavenumber dynamics of a defect laden solution of \eqref{e:cgl-c} which connects stripe solutions with small transverse wavenumber $k_{y,+} = \delta q_+$ for $y>0$ and $k_{y,-} = \delta q_-$ for $y>0$, with $0<\delta\ll1$ and $q_\pm = \mc{O}(1)$. In particular, Fig. \ref{f:sink-source} depicts a defect with $q_- = 3$ and $q_+=1$. Such a defect solution was obtained numerically with an initial condition of the form 
\begin{align}\label{e:cglsrc-sk}
A(\xi,y,t) &= h(-\xi)\Big(h(y)r_-\re^{\ri(k_{x,-}\xi+ k_{y,-}y)} \notag\\
&+ h(-y)r_+\re^{\ri(k_{x,+}\xi + k_{y,+}y)} \Big), \quad
\end{align}
where $h(z)$ denotes the Heaviside function, $r_\pm^2 =\sqrt{1- (k_{x,\pm}^2 + k_{y,\pm}^2)}$, and the wavenumbers $k_{x,\pm}$ are chosen so that $k_{x,\pm} = k_{x,\rf}(k_{y,\pm})$, using the computed wavenumber curves depicted in Fig. \ref{f:auto}.   
Since the numerical computation uses periodic boundary conditions in the vertical direction, this solution consists of two well-separated defects, one a source (left in Fig. \ref{f:sink-source}b) and one a sink (right in Fig. \ref{f:sink-source}b). We remark that the sink creates a grain boundary, also known as a domain wall, in the far-field while the source creates a wavenumber fan between the two striped states.  



\begin{figure*}
\centering
\hspace{-0.15in}(a)\hspace{-0.04in}\includegraphics[trim = 0.05cm 0.0cm 0.05cm 0.0cm,clip,width=0.33\textwidth]{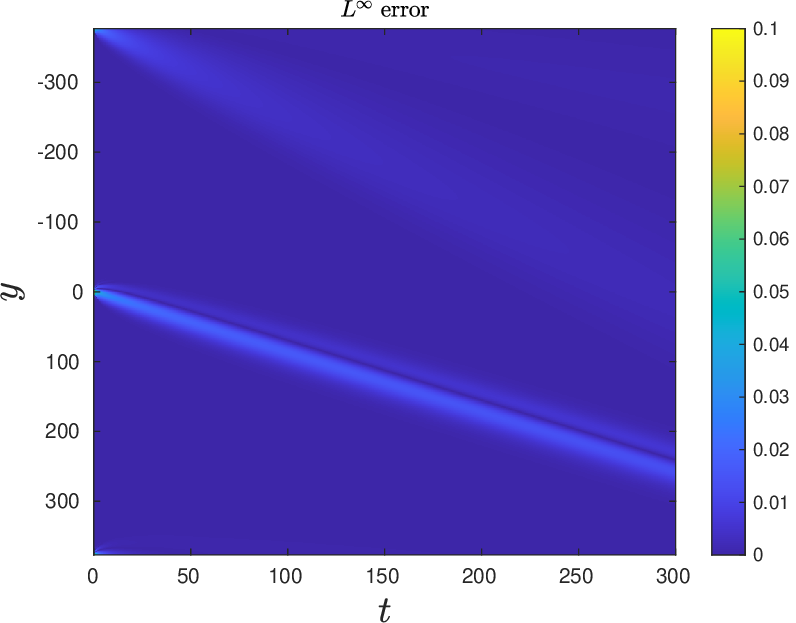}\hspace{-0.0in}
(b)\hspace{-0.04in}\includegraphics[trim = 0.0cm 0.0cm 0.05cm 0.0cm,clip,width=0.31\textwidth]{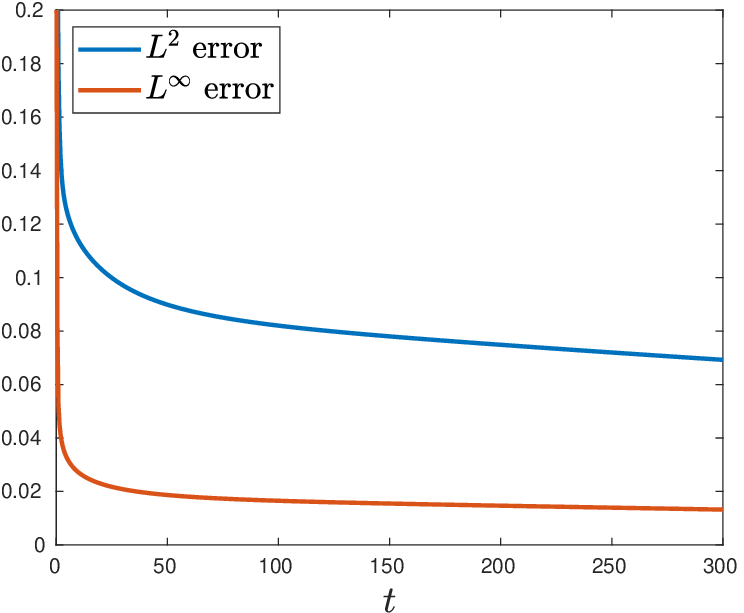}\hspace{-0.05in}
(c)\hspace{-0.04in}\includegraphics[trim = 0.05cm 0.0cm 0.05cm 0.0cm,clip,width=0.33\textwidth]{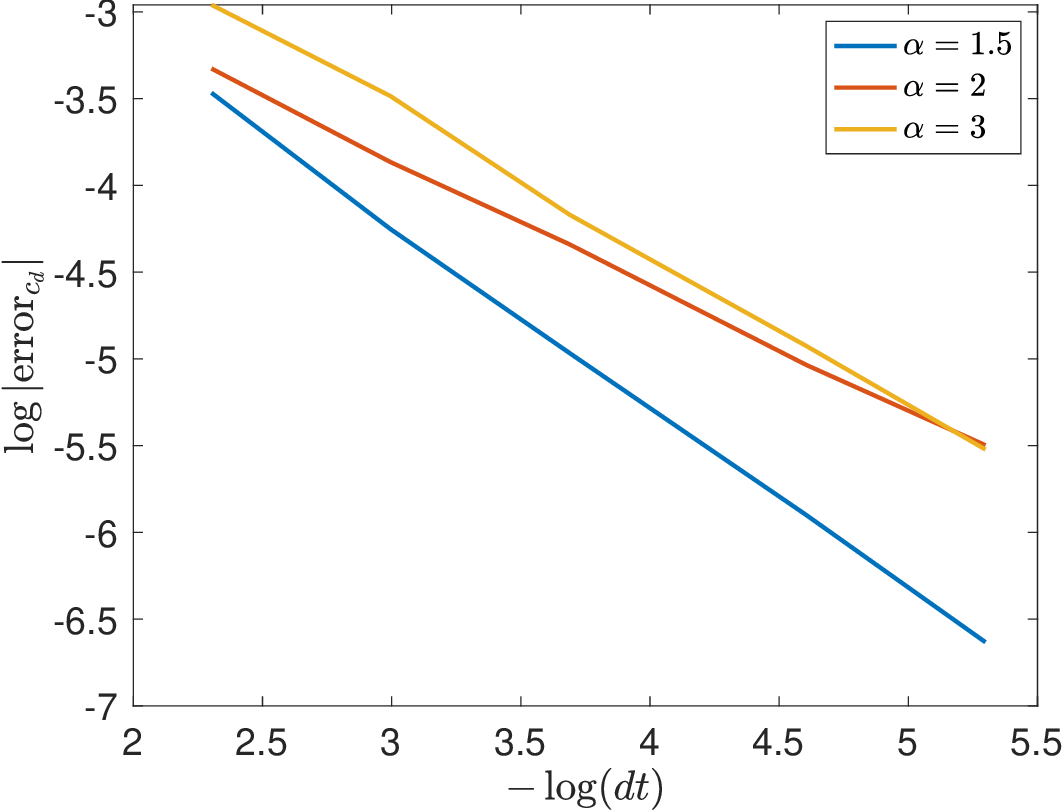}\hspace{-0.3in}
\caption{ (a): Spacetime diagram of the absolute error between measured transverse wavenumber $\mathrm{Im}\, \partial_y A/A$ and the rescaled viscous Burgers' solution $\delta \Psi(\delta y, \delta^2 t)$, for the same initial condition and parameters as Fig. \ref{f:sink-source}; (b): Plots of the corresponding $L^2$ and $L^\infty$ norm in $y$, note the large error for $t\sim0$ is due to Gibbs oscillations in the measured wavenumber; (c): Plot of the $dt$-convergence of error between the numerically measured defect speed and the theoretical prediction $c_d = c_{g,0} + \delta c_*$, for range of $\alpha$ values.    }\label{f:shock}
\end{figure*}

\begin{figure*}
\centering
\hspace{-0.15in}(a) \hspace{-0.02in}\includegraphics[trim = 0.05cm 0.0cm 0.05cm 0.05cm,clip,width=0.3\textwidth]{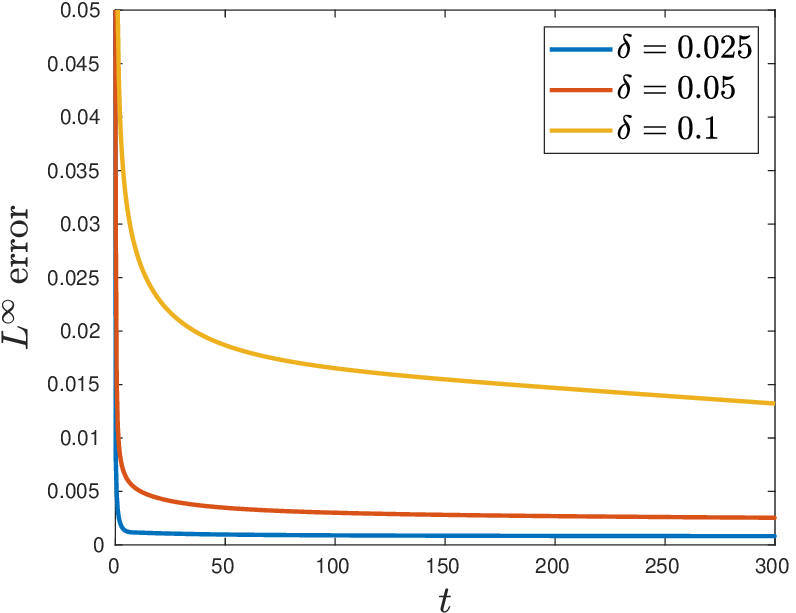}\hspace{-0.0in}
(b) \hspace{-0.05in}\includegraphics[trim = 0.0cm 0.0cm 0.05cm 0.05cm,clip,width=0.3\textwidth]{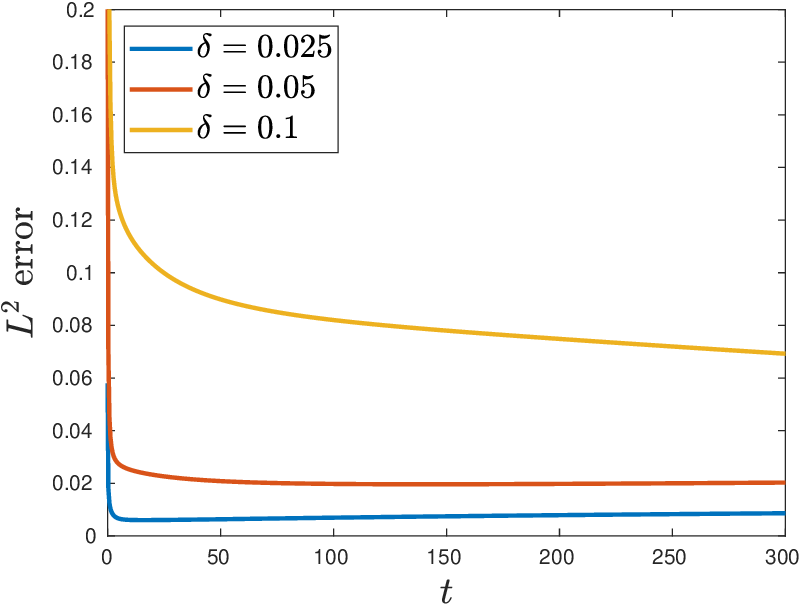}\hspace{-0.05in}
\caption{Evolution of (a) the $L^2$ and (b) the $L^\infty$ error, between the measured transverse wavenumber in \eqref{e:cgl-c} and the rescaled viscous Burgers' solution \eqref{e:vb}, for the source-sink initial data \eqref{e:cglsrc-sk} for $\delta = 0.025,0.05,0.1$, showing decrease in the error as $\delta\rightarrow0$. Same initial condition and parameters as Fig. \ref{f:sink-source}.  }\label{f:l2li_err}
\end{figure*}

As the periodic wavetrains are relative equilibria, the local transverse wavenumber of the defect may be numerically measured as
\beq\label{e:psimeas}
\psi(y,t) = \mathrm{Im}\, A_y(\xi_0,y,t)/A(\xi_0,y,t).
\eeq
To determine the coefficient $\omega_\rf''(0)$ of the viscous Burgers equation, we measure $k_{x,\rf}(0)$ and $k_{x,\rf}''(0)$ using the numerically computed curve $k_{x,\rf}(k_y)$ depicted in Figure \ref{f:auto}b. For the second-derivative, we performed a quadratic fit of the data near the origin $k_y = 0$. We then compared the measured wavenumber $\psi(y,t)$ to the predicted transverse wavenumber dynamics coming from the Burgers' equation. We use the initial local wavenumber $\psi(y,0)$ as the initial data for the viscous Burgers' equation \eqref{e:vb}, 
$$
\Psi(Y,0) = \psi(Y/\delta,0)/\delta. 
$$
For the specific initial condition \eqref{e:cglsrc-sk}, $\Psi(Y,0)  = q_-$ for $Y\in[-\delta L_y/2,0]$ and $\Psi(Y,0) = q_+$ for $Y\in[0,\delta L_y/2)$. 
We then numerically integrate the viscous Burgers' equation, posed on the scaled periodic domain, forward in time and then scale back to obtain a prediction for the transverse wavenumber dynamics at time $t>0$,
$$
\psi(y,t) \approx \delta \Psi(\delta y,\delta^2 t).
$$
As depicted in Figure \ref{f:sink-source}, we find good agreement between the prediction from the Burgers' equation \eqref{e:vb} and the numerically measured wavenumber.  See Figure \ref{f:shock}a-b for the evolution of the error. Figure \ref{f:l2li_err} depicts errors for the wavenumber prediction for initial conditions \eqref{e:cglsrc-sk} for a range of $\delta$ values, showing that both the $L^2$ and $L^\infty$ norm (in $y$) decreases as $\delta\rightarrow0$ and the temporal regime of validity appears to scale like $\delta^{-2}$ (roughly consistent with the results of (Ref. \onlinecite{dmwt})). 
We also note that the initial large error and sharp decrease is due to numerical instabilities in the measured wavenumber $\psi$. In particular, the derivative $A_y$, used in the measured wavenumber \eqref{e:psimeas}, is calculated spectrally so that the sharp jump at $y\sim0$ in \eqref{e:cglsrc-sk} induces Gibbs-type oscillations. Parabolic regularization in \eqref{e:cgl} smooths the instantaneous jump in wavenumber, leading to a local decrease in the error for short times. 

We also note that the viscous Burgers' modulation equation gives accurate predictions of the defect speed, $c_\mathrm{d}$, in the transverse direction.
For example, considering the right-ward traveling defect where $q_->q_+>0$, we use the fact that $\omega_\rf''(0)>0$ to approximate the transverse group velocities, $c_\mathrm{g,\pm}:= \omega_\rf'(k_{y,\pm}) \approx c_{g,0}+\omega_\rf''(0) \delta q_\pm$, of the asympototic wave trains,
\begin{eqnarray}
c_\mathrm{g,-} > c_\mathrm{d}>c_\mathrm{g,+},
\end{eqnarray}
so that they point inwards and the defect behaves as a sink. 
The sink defect portion of the solution in Figure \ref{f:sink-source} corresponds to a traveling shock wave solution $q_*(Y - c_* T)$ of \eqref{e:vb} with speed $c_*$ connecting the asymptotic states $q_\pm$ at $Y = \pm\infty$ respectively. The Lax entropy condition for a traveling shock requires, $\omega_\rf''(0)(q_--q_+)>0$. The Rankine-Hugoniot criterion in this case gives the shock speed as $c_* = (\omega_\rf''(0)q_- + \omega_\rf''(0)q_+)/2$. 
This allows us to obtain a prediction for the transverse defect speed in the full 2-D system. Since the parallel stripe has transverse group velocity $c_{g,0} := \omega_\rf'(0)  = 0$, we find the defect speed to be 
\begin{eqnarray}\label{e:vb-sp}
&c_\mathrm{d} = c_{g,0} + \delta c_* = \delta \frac{\omega_\rf''(0)}{2}(q_- + q_+).
\end{eqnarray}
See also Sec. 1.3 of (Ref. \onlinecite{dmwt}) for more detail on such calculations. Figure \ref{f:shock}c shows good agreement of the numerically measured defect speed with that predicted by the modulation equation. In particular the measured shock speed converges to the predicted speed with rate $\mathcal{O}(dt)$ as $dt$ is reduced. We also observe that source defects with $ \omega_\rf'(k_y^-)< c_\mathrm{d} < \omega_\rf'(k_y^+)$ behave as rarefaction waves.

\begin{figure*}
\centering
\hspace{-0.15in}(a)\hspace{-0.05in}  \includegraphics[trim = 0.cm 0.cm 0.cm 0.cm,clip,width=0.3\textwidth]{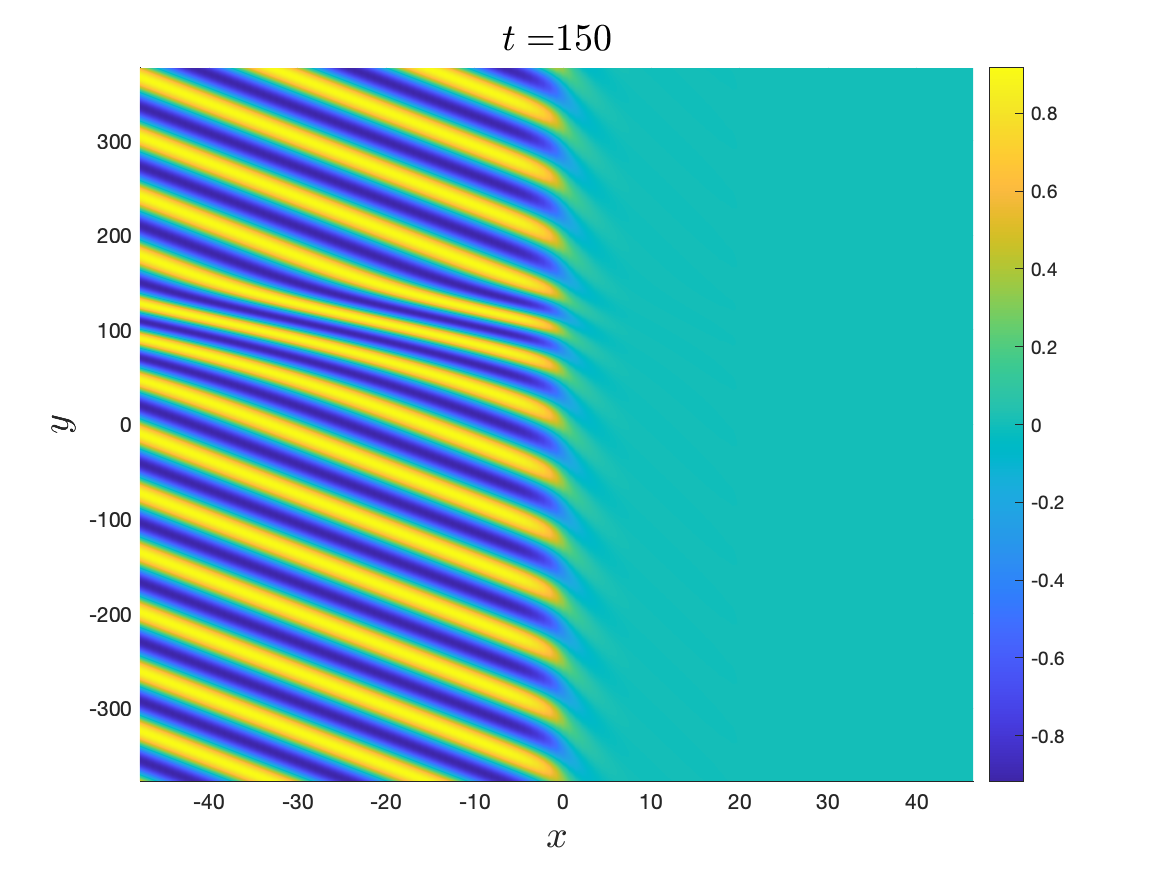}\hspace{-0.2in}
(b)\hspace{-0.05in}  \includegraphics[trim = 0.cm 0.cm 0.cm 0.cm,clip,width=0.27\textwidth]{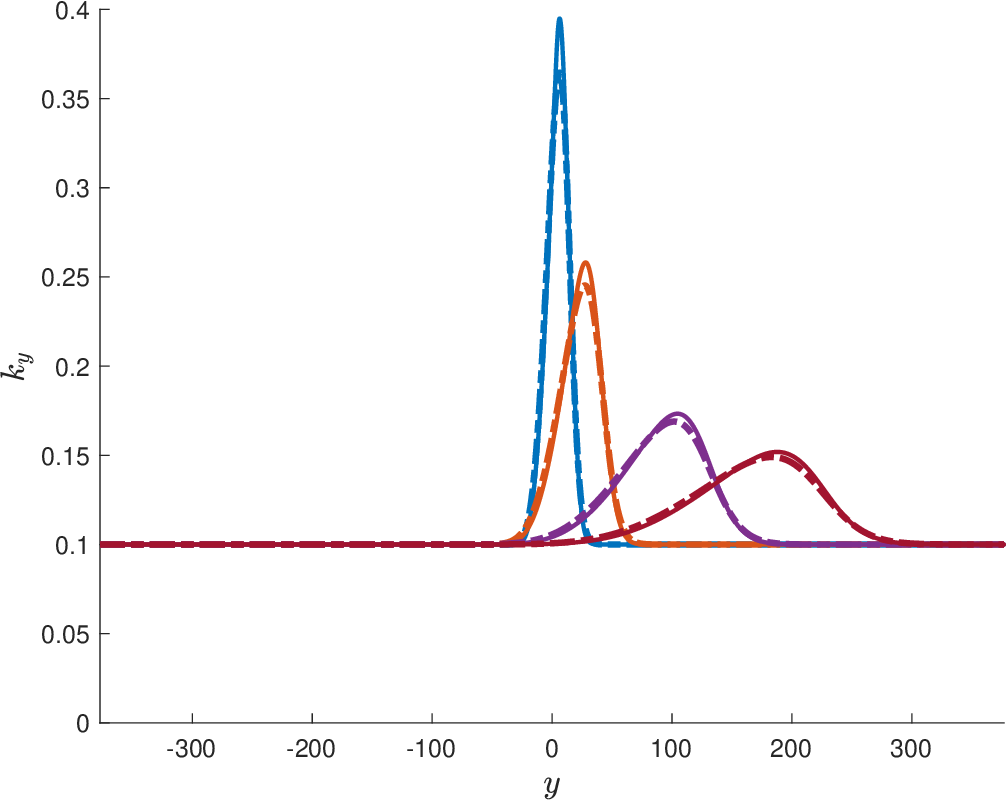}\hspace{-0.1in}
(c) \hspace{-0.05in} \includegraphics[trim = 0.cm 0.cm 0.cm 0.cm,clip,width=0.27\textwidth]{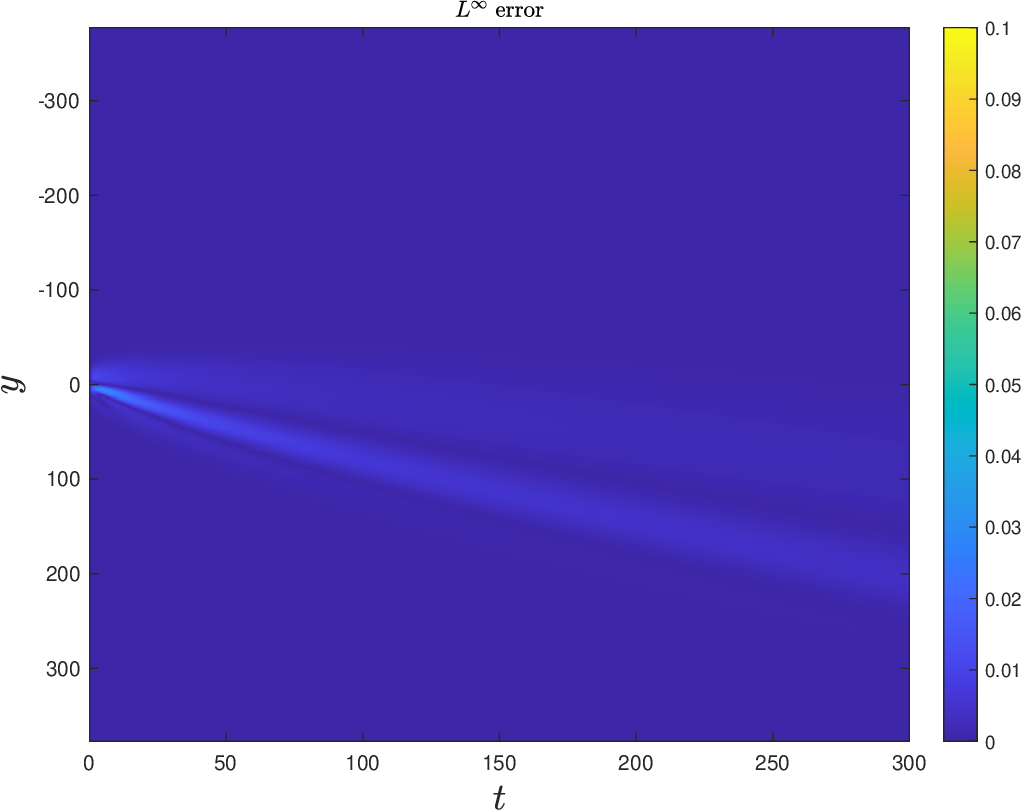}\hspace{-0.2in}\\
(d) \hspace{-0.05in} \includegraphics[trim = 0.cm 0.cm 0.cm 0.cm,clip,width=0.27\textwidth]{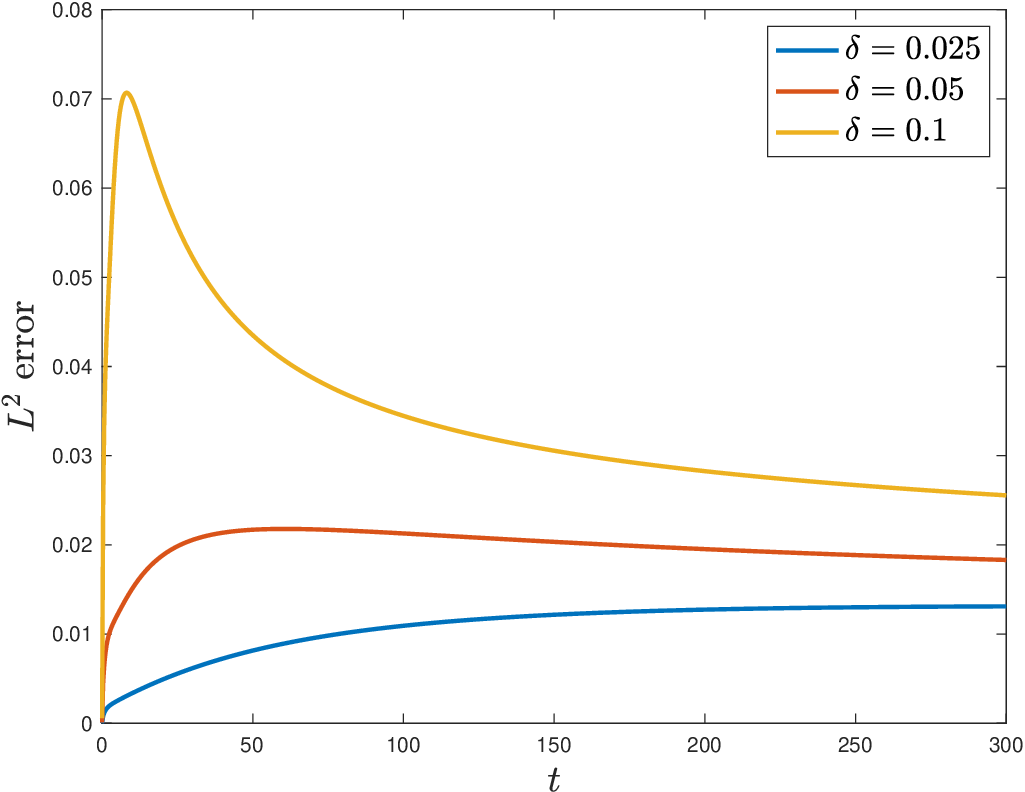}\hspace{-0.0in}
(e) \hspace{-0.05in} \includegraphics[trim = 0.cm 0.cm 0.cm 0.cm,clip,width=0.27\textwidth]{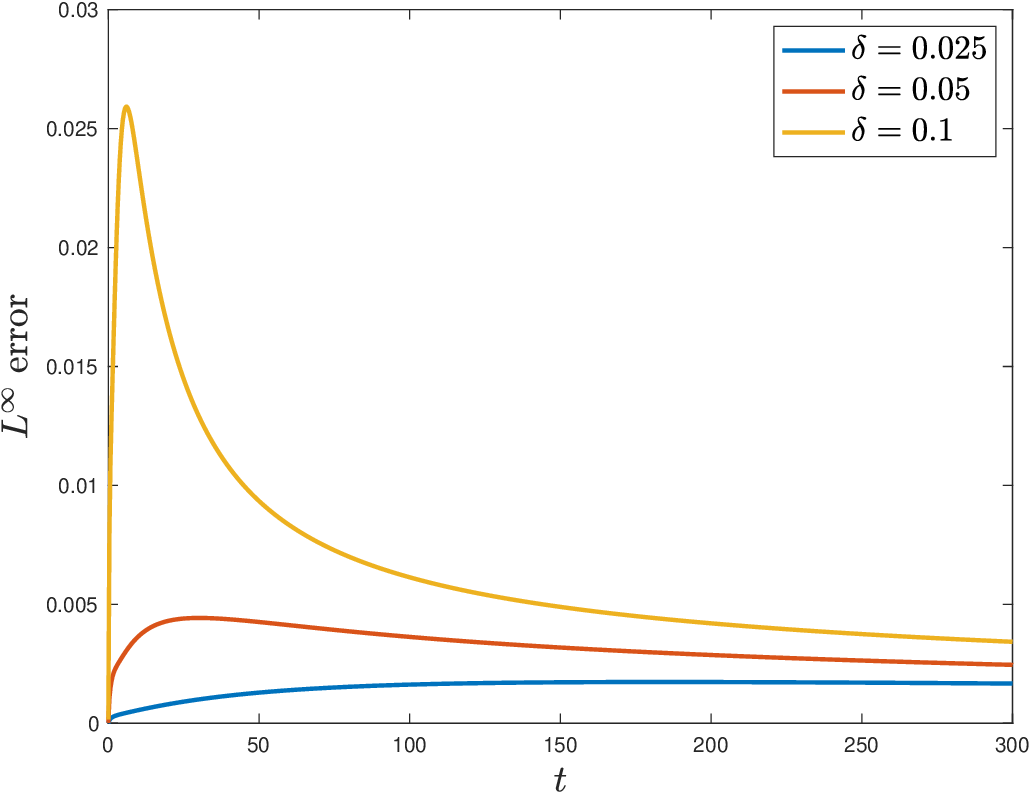}\hspace{-0.2in}
\caption{Top row: Localized phase-slip defect solution of \eqref{e:cgl-c} with initial data \eqref{e:loc_in}, $c = 2.5, \alpha = 3, \gamma = 1$ and $\delta = 0.1$, numerical time step $dt = 0.0025$. (a): plot of $\mathrm{Re}\, A(\xi,y,t = 150)$; (b): Comparison of numerically measured transverse wavenumber (solid line) with the associated prediction from viscous Burgers equation \eqref{e:vb}  (dashed line) for times $t = 5 \text{(blue)}, 30 \text{(orange)}, 150 \text{(purple)}, 300 \text{(dark red)}$; (c): point-wise absolute error between measured and predicted wavenumber; Bottom row: plots of both the $L^2$ and $L^\infty$ errors, (d) and (e) respectively, for the localized phase-slip for $\delta = 0.025,0.05,0.1$.}\label{f:cgl_loc}
\end{figure*}

\subsubsection{Phase-slip defect modulation }

  As another example, we consider a localized wavenumber perturbation of a quenched stripe in \eqref{e:cgl-c}. Figure \ref{f:cgl_loc} depicts the transverse dynamics of this localized defect, initiated by an initial condition of the form
\begin{align}
A(\xi,y,t) &= h(-\xi) \sqrt{1 - k^2} \mathrm{exp}\left[ \ri (k_x \xi + \delta y + \phi_0(\delta y)) \right],\quad \label{e:loc_in}
\end{align}
where $\phi_0(Y) = \pi \mathrm{erf}(Y)$ and $\mathrm{erf}(Y) = 2\pi^{-1/2} \int_0^Y \re^{-t^2} dt$ denotes the Error function. This initial condition induces a phase slip perturbation which does not alter the asymptotic wavenumber for $|y|$ large. We chose transverse wavenumber $k_y = \delta$ so that the corresponding scaled wavenumber profile satisfies $\Psi(Y) =1+  4\pi^{1/2} \re^{-Y^2}.$ We once again find good agreement between the transverse wavenumber dynamics and the viscous Burgers prediction, with both $L^2$ and $L^\infty$ error decreasing as $\delta$ is decreased; see Figure \ref{f:cgl_loc}(c-e).

\section{Modulations in the quenched Swift-Hohenberg equation}
\begin{figure}[h!]
\centering 
(a) \includegraphics[trim = 0.0cm 0.cm 0.0cm 0.0cm,clip,width=0.3\textwidth]{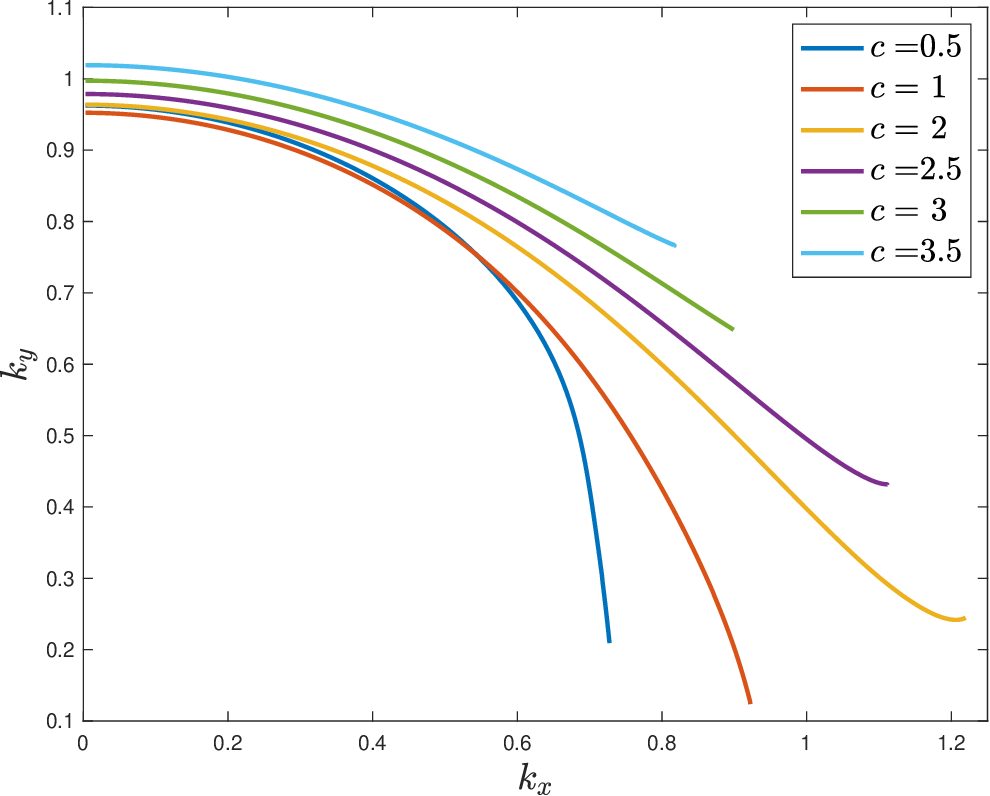}\hspace{-0.0in}\\
(b) \includegraphics[trim = 0.0cm 0.cm 0.0cm 0.0cm,clip,width=0.3\textwidth]{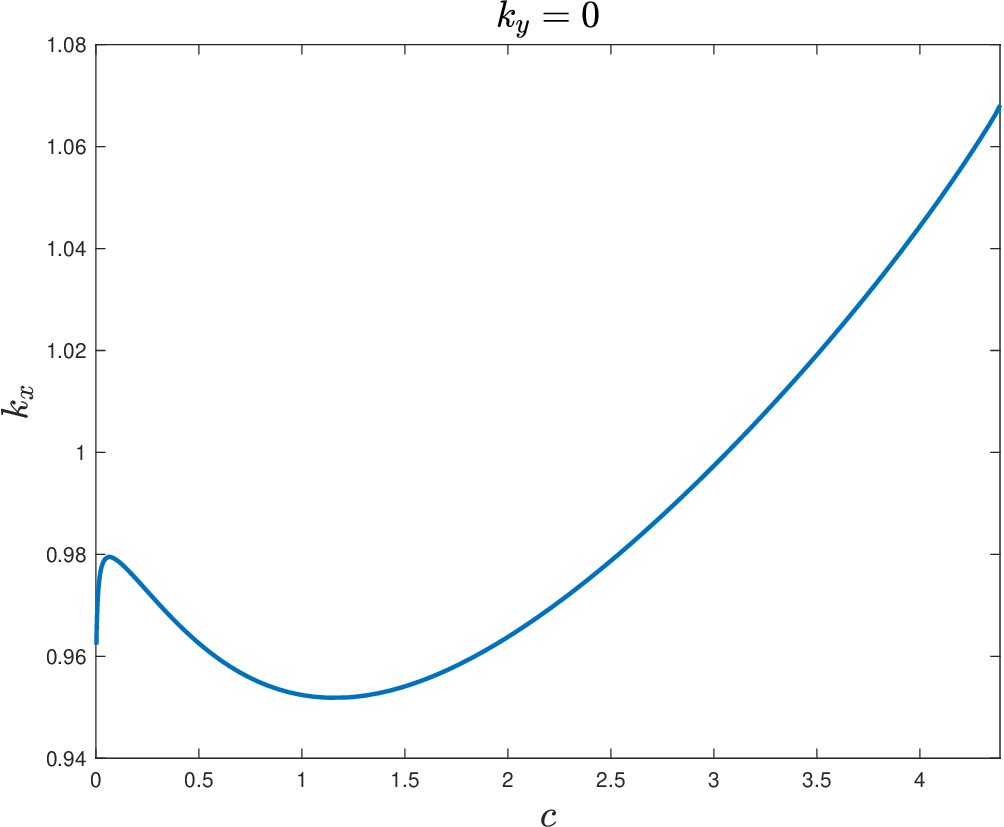}\hspace{-0.2in}\\
(c) \includegraphics[trim = 1.75cm 0.cm 0.25cm 0.0cm,clip,width=0.4\textwidth]{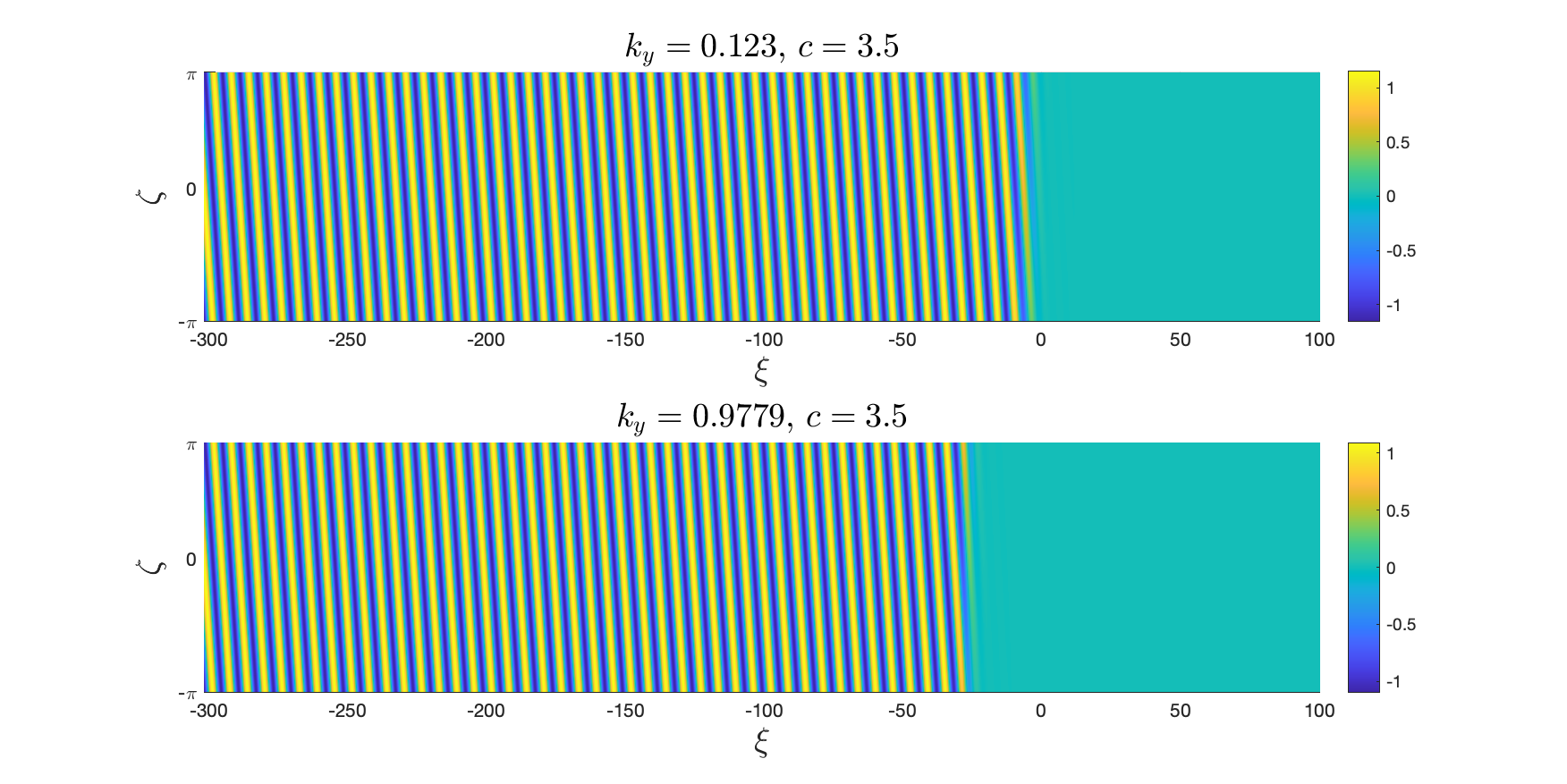}\hspace{-0.2in}
\caption{Swift-Hohenberg wavenumber selection curves for $\mu = 1$, depicting $k_{x,\rf}(c,k_y)$ for (a) $c$ fixed, and $k_y$ varied, and (b) $k_y = 0$ fixed and $c$ varied; (c): Example front solution profiles $u_\rf(\xi,\zeta)$ with $c = 3.5$ and $k_y = 0.123, 0.9779$; $\mu = 1$ throughout.}\label{f:sh-kx}
\end{figure}

To show the applicability of this approach, we also employ it to describe stripe modulations in the quenched Swift-Hohenberg (SH) equation (Ref. \onlinecite{swift1977hydrodynamic}) with supercritical nonlinearity,
\begin{eqnarray}\label{e:sh00}
u_t  &= -(1+\partial_x^2+\partial_y^2)^2u  + \mu\chi(x - ct) u - u^3,\\
& \qquad (x,y)\in\R^2, \quad \mu>0.\notag
\end{eqnarray}
We choose this equation  as it does not have exact closed form periodic stripe solutions - only leading-order expansions at onset $0<\mu\ll1$ -  and thus one generally must compute the viscous Burgers' coefficients, $\lambda_\rlin''(0)$ and $\omega_\rf''(0)$ with asymptotic expansions or numerically.  
For $\chi\equiv1$ and $\mu>0$, it is well-known that \eqref{e:sh00} has stable stripe equilibrium solutions $u_p(k_x x + k_y y;k)$, $2\pi$-periodic in the first argument and dependent only on the bulk wavenumber $k^2 = k_x^2+k_y^2$ due to the rotational invariance of the homogeneous system. The range of $k$ values for which stripes exist and are stable is determined by the Busse balloon (Ref. \onlinecite{mielke97}). To study quenched fronts, we once again move into the co-moving frame $\xi = x- ct$,
\begin{equation}\label{e:sh-c}
u_t  = -(1+\partial_\xi^2+\partial_y^2)^2u  + c\p_\xi u + \mu\chi(\xi) u - u^3.
\end{equation}
Previous works (Ref. \onlinecite{gs3,MR3958766,goh2023growing}) have studied front solutions of this equation of the form $u_\mathrm{f}(\xi,k_y y + \omega t)$, periodic in the second variable $\zeta = k_y y + \omega t$, which satisfy
\begin{align}
0&= -(1+\partial_\xi^2+k_y^2\partial_\zeta^2)^2 u_\rf + (c\partial_\xi- \omega\partial_\zeta) u_\rf + \mu\chi(\xi)u_\rf - u_\rf^3,\qquad\quad \label{e:shq}\\
0&= \lim_{\xi\rightarrow-\infty} u_\rf(\xi,\zeta),\quad 0= \lim_{\xi\rightarrow-\infty} u_\rf(\xi,\zeta) - u_p( k_x\xi + \zeta;k),\notag\\
& u_\rf(\xi,\zeta) = u_\rf(\xi,\zeta+2\pi).
\end{align}
We note that in this co-moving, co-rotating frame, under the 1:1 resonance condition $\omega = c k_x$, stripe equilibria $u_p$ become $2\pi$-periodic in $\zeta$, with $k_x x + k_y y = k_x \xi + \zeta$. As in \eqref{e:cgl-c}, the horizontal wavenumber $k_{x,\rf}$ of quenched front solutions, $u_\rf$, is generically selected by parameters $(c,k_y)$ and can be written locally as a graph over these two variables. We  note the works (Refs. \onlinecite{MR3958766,goh2023growing}) gave a near complete numerical description of the moduli space of patterns using a far-field core numerical continuation approach.  See Figure \ref{f:sh-kx} for depictions of select wavenumber selection curves for $c$ fixed and $k_y$ varying, and vice-versa.

 As before, since we shall fix $c>0$, we denote these fronts as $u_\mathrm{f}(\xi,\zeta;k_y)$, and the corresponding selected wavenumber as $k_{x,\mathrm{f}}(k_y)$. 
  The work (Ref. \onlinecite{gs3}) showed the existence of parallel stripes with $k_y = 0$ for $0<\mu\ll1$ and in the fast quench regime where $c\lesssim c_\mathrm{lin}$ using center manifold techniques. It also used functional analytic techniques to prove that for any $c\in (0,c_\rlin)$ fixed,  parallel striped fronts perturb smoothly in $k_y\sim 0$, and the horizontal wavenumber satisfies the expansion 
\begin{align}
&k_{x,\mathrm{f}}(k_y) = k_{x,\mathrm{f}}(0) + \beta_2 k_y^2 + \mathcal{O}(k_y^4), \label{e:shkxex}\\
& \quad\qquad \beta_2 = \frac{1}{c}\langle 2\p_\zeta^2 (1+\partial_\xi^2) u_\mathrm{f}(\cdot,\cdot;0)\,,e_* \rangle_{L^2_\eta},
\end{align}
where $\langle v_1,v_2\rangle_{L^2_\eta} = \int_{\R\times[0,2\pi)} \re^{2\eta|\xi|} v_1\bar v_2 d\xi d\zeta$ denotes the exponentially-weighted $L^2$ inner product, and $e_*$ spans the kernel of the $L^2_\eta$-adjoint, ${\mb{L}}^*$, of the linearization of \eqref{e:shq}  about the front
\begin{eqnarray*}
&\mb{L}v:= -\omega v_\zeta + cv_\xi - (1+\partial_\xi^2)^2 v + \mu\chi v - 3 (u_\mathrm{f}(\cdot,\cdot;0))^2 v.
\end{eqnarray*}
 We now briefly discuss how the parameters $\lambda_\rlin''(0)$ and $\omega_\rf''(0)$ can be computed. The derivation ofthe modulation equation follows a similar line as for CGL, which is described in Appendix \ref{a:1}.

\subsection{Nonlinear Burgers parameter}
For $\omega_\rf''(0)$, we use the 1:1 resonance condition and the $k_y$ expansion \eqref{e:shkxex} above to obtain
\beq
\omega_\rf(k_y) = c k_{x,\rf}(k_y) = c \left(  k_{x,\mathrm{f}}(0) + \beta_2 k_y^2 + \mathcal{O}(k_y^4) \right).
\eeq
This readily gives
\beq\label{e:shom2}
\omega_\rf'(0) = 0 , \qquad \omega_\rf''(0) = 2\beta_2 c
\eeq
We note that this quantity could also be obtained by differentiating the front equation twice in $k_y$, evaluating at $k_y = 0$ and then projecting onto the adjoint kernel element $e_*$ defined above. In practice, we calculate $\omega_\rf''(0)$ by once again performing a quadratic fit of the numerical continuation data (Fig. \ref{f:sh-kx}a) for the curve $k_{x,\rf}(k_y)$ for $k_y\sim0$, obtaining a numerical prediction of the quadratic coefficient $\beta_2$.  Numerical computations of these curves (see Fig. \ref{f:sh-kx} as well as Refs. \onlinecite{avery2019growing,goh2023growing})  indicates that $\omega_\rf''(0)<0$ for a wide range of $c$ values except possibly for $0<c\ll1$.

\subsection{Effective Diffusivity}\label{ss:sh_effdiff_sh}
The effective diffusivity, $\lambda_\rlin''(0)$, for $y$ perturbations of parallel striped fronts just behind the quenching line can be obtained using a Fourier-Bloch wave analysis (Refs. \onlinecite{mielke97,gs3}). To summarize, one sets $k_y = 0$ and linearizes \eqref{e:sh-c} about the front $u_\rf$. Since the far-field state at $\xi = +\infty$ is exponentially stable, it suffices to consider the linearization at $\xi = -\infty$, that is linearizing about the asymptotic roll state $u_p(k_x\xi + \tau; k_x)$ with $\chi \equiv \mu$,
\begin{eqnarray}
&\mathbb{L}_pw:=-(1+\Delta)^2w+\mu w+c_x\p_\xi w\notag\\
&\qquad\qquad-3u_p^2(k_x\xi+\tau;k_x)w-\omega\p_\tau w.
\end{eqnarray}
The $L^2$ spectrum of this operator can be studied using the Floquet-Fourier-Bloch ansatz 
\begin{eqnarray}\label{e:bloch}
    &w(\xi,y,t)=e^{im\tau}e^{iny}e^{\nu\xi}b(k_x\xi+\tau;m,n,\nu),\\
    &\qquad m\in\Z,n\in\R,\nu\in \ri[0,2\pi/k_x).\notag
\end{eqnarray}
for $b(z;m,n,\nu)$ a $2\pi$-periodic function in $z$. Inserting this into $\mathbb{L}_p w = \lambda w$, and using the fact that $\omega = ck_x$, we obtain a family of eigenvalue problems 
\begin{widetext}
\beq\label{e:bl-eval}
\lambda b = \mathcal{B}(m,n,\nu)b:=-(1+(k_x\p_z+\nu)^2-n^2)^2b+\mu b+(c_x \nu -\omega\ri m) b-3u_p(z;k_x)^2b.
\eeq
\end{widetext}
Evaluating \eqref{e:bl-eval} at $m=0,n=0,\nu=0$, we find that $b=\p_zu_p$ is an eigenfunction with eigenvalue $\lambda=0$. A perturbative approach then gives a family of eigenvalue-eigenfunction pairs $(\lambda_\rlin(\nu),w_\rlin(\nu))$ emanating from $(0,\p_zu_p)$. Since $\mathcal{B}(0,0,0)$ is $L^2([0,{2\pi}))$-self-adjoint with $\mathrm{ker}\,  \mathcal{B}(0,0,0) = \mathrm{span}\, \p_z u_p$, we also have that $\p_z u_p \in \mathrm{Rg}\, \mathcal{B}(0,0,0)^\perp$. Letting $b_*$ be a scalar multiple of $\p_z u_p$ so that $\langle b_*,  b_* \rangle_{L^2(0,2\pi)} = 1,$ and differentiating the eigenvalue equation \eqref{e:bl-eval} in $\nu$, we obtain 
\beq
\lambda''_\mathrm{lin}(0) = \langle4(1+(k_x\p_z)^2)\p_zu_p,b_*\rangle_{L^2(\mathbb{T}_{2\pi})}.
\eeq
We computed this inner product numerically using Newton's method to solve a finite difference discretization of the periodic boundary value problem $0 = -(1+k^2 \partial^2_\theta)^2 u_p + \mu u_p - u^2$,  and an iterative linear solver to obtain a numerical discretization of the kernel element $b_*$.

\begin{figure}[h!]
\centering
\hspace{-0.15in} (a) \hspace{-0.04in} \includegraphics[trim = 0.0cm 0.cm 0cm 0.cm,clip,width=0.33\textwidth]{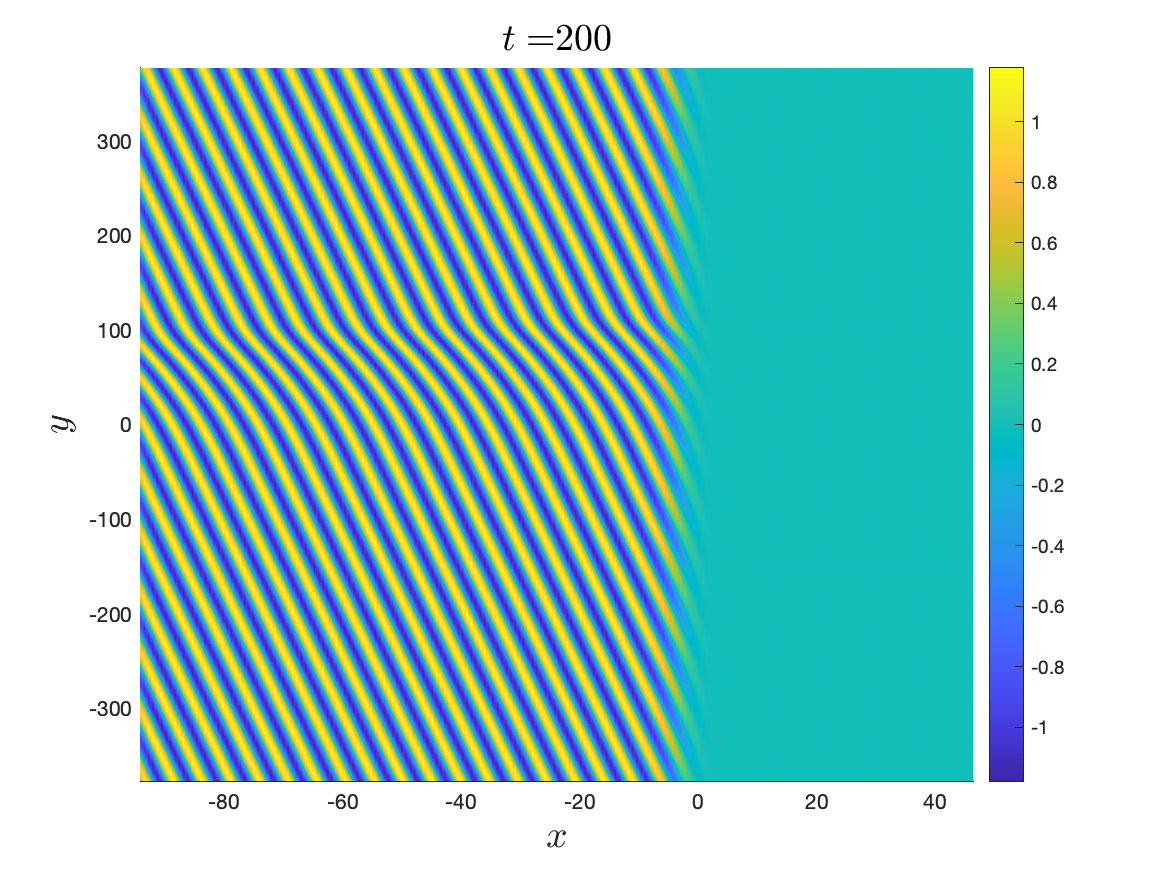}\hspace{-0.2in}\\
\noindent\hspace{-0.15in} (b) \hspace{-0.04in}\includegraphics[trim = 0.0cm 0.cm 0cm 0.cm,clip,width=0.30\textwidth]{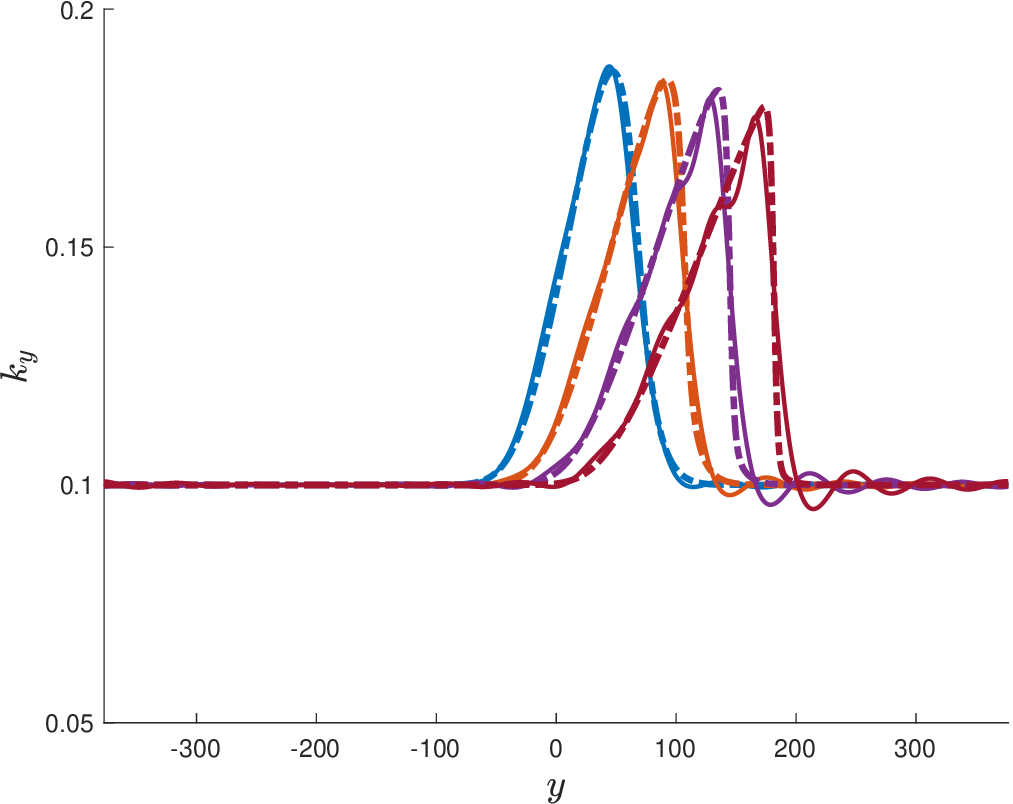}\hspace{-0.2in}\\
\noindent \hspace{-0.15in} (c) \hspace{-0.04in} \includegraphics[trim =0.0cm 0.cm 0cm 0.cm,clip,width=0.30\textwidth]{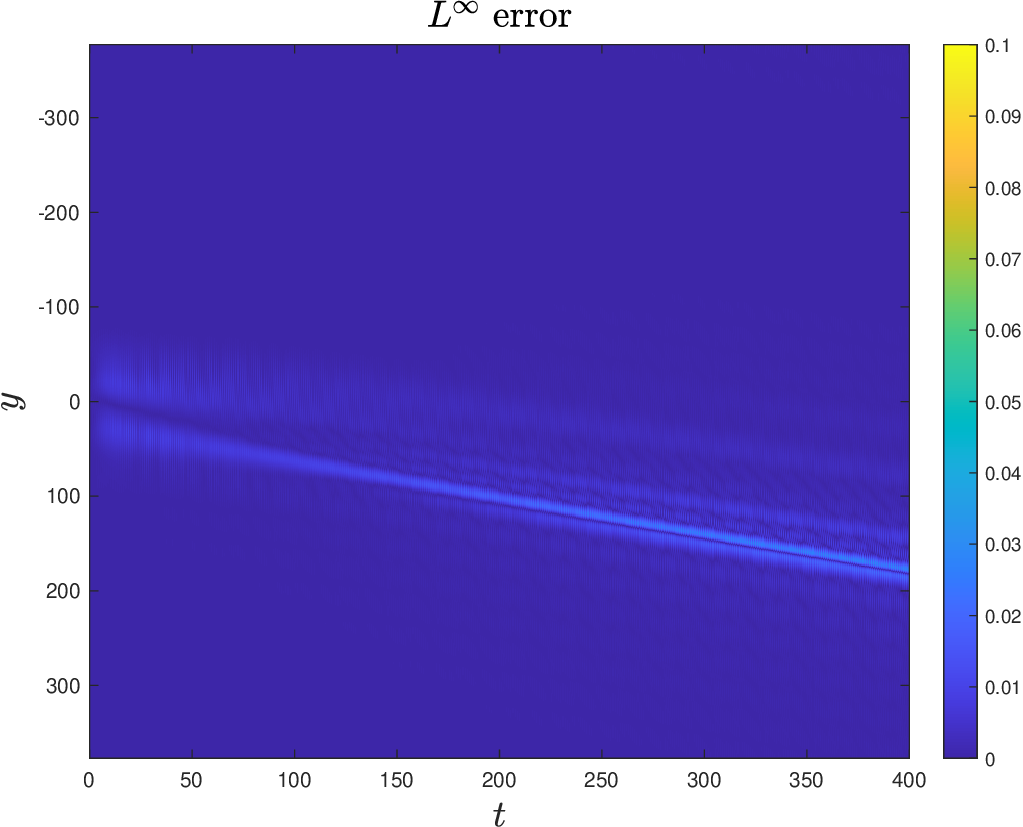}\hspace{-0.2in}
\caption{Localized phase perturbation in equation \eqref{e:sh-c} of a striped front with $(k_x,k_y) = (0.993,0.1)$; Initial condition $u(\xi,y,0) = \sqrt{4\mu/3} \cos(k_x x + k_y y + \phi_0(\delta y)) h(-\xi)$ (similar to \eqref{e:loc_in}), with $c = 3,\mu = 1$; the Burger's parameters were computed numerically as $\lambda_\rlin''(0) = 0.3202...$, $\omega_\rf''(0) = -2.67471... $; (a): solution profile $u(\xi,y,t=200)$, (b): Transverse wavenumber dynamics at $\xi = \xi_0 = -1$ fixed (solid) plotted against the rescaled viscous Burgers' solution \eqref{e:vb} at times $t = 100 \text{(blue)},200\text{(orange)},300\text{(purple)},400\text{(red)};$ (c): plot of absolute error between measured and predicted wavenumber profiles. }\label{f:sh_mod_num} 
\end{figure} 
\begin{figure*}
\centering
\hspace{-0.15in}  (a) \hspace{-0.04in}\includegraphics[trim = 0.0cm 0.cm 0cm 0.cm,clip,width=0.3\textwidth]{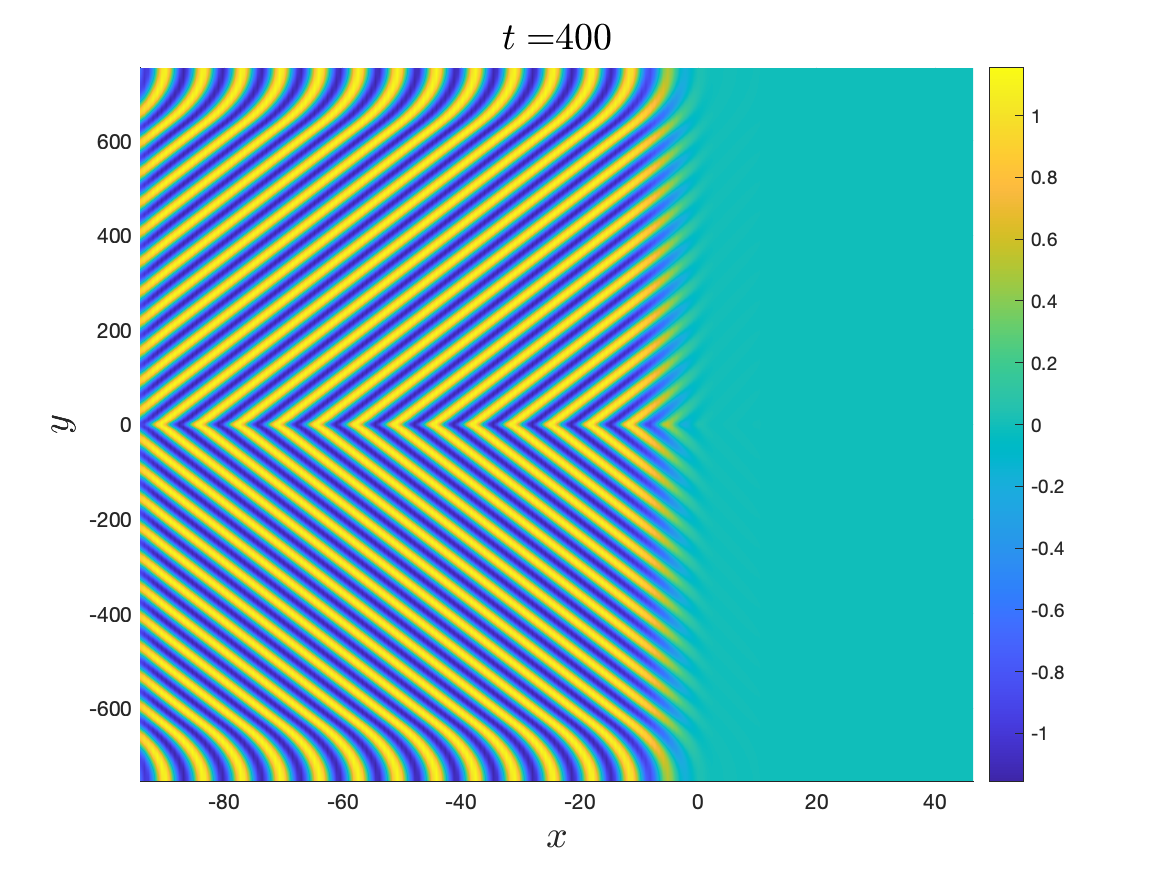}\hspace{-0.2in}
 (b) \hspace{-0.04in}\includegraphics[trim = 0.0cm 0.cm 0cm 0.cm,clip,width=0.3\textwidth]{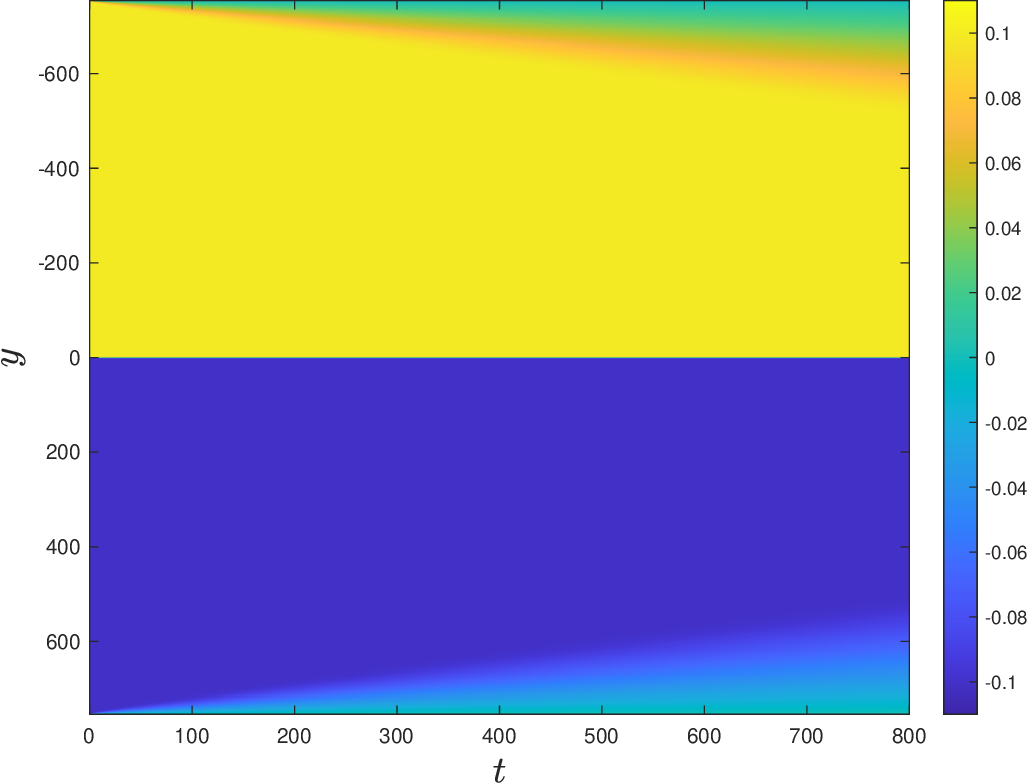}\hspace{-0.05in}
 (c) \hspace{-0.04in}\includegraphics[trim = 0.0cm 0.cm 0cm 0.cm,clip,width=0.3\textwidth]{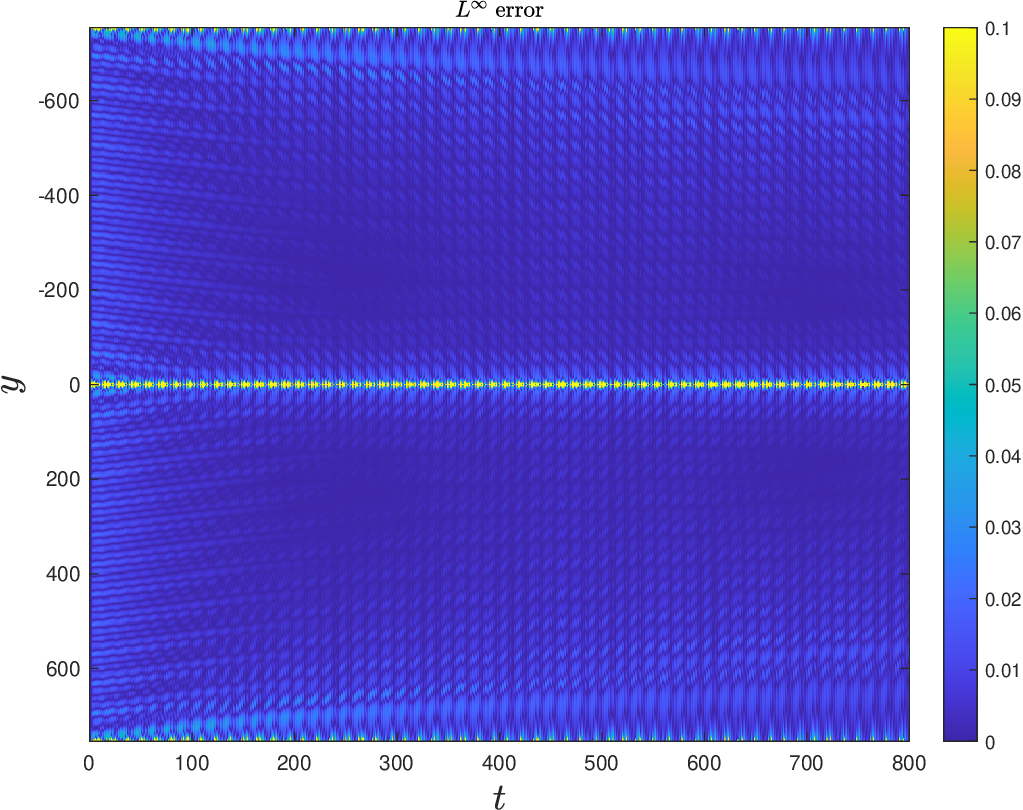}\hspace{-0.2in}
\caption{Grain boundary defect in equation \eqref{e:sh-c} with $c = 3,\mu = 1$, and Burger's parameters as in Fig. \ref{f:sh_mod_num}; (a): snapshot of grain boundary type solution of \eqref{e:sh-c} with $\mu = 1$, $c = 3,$ and transverse wavenumbers $k_y = \pm0.1$, plotted at $t = 200$, with convex grain boundary at center of domain $y\sim0$, and concave grain boundary at the boundary of the (periodic) computational domain. Center: space-time diagram of corresponding rescaled Burgers' solution $\delta\Psi(\delta y,\delta^2 t)$ of \eqref{e:vb}; Right: spacetime diagram of absolute error between measured and predicted wavenumbers. }\label{f:sh_mod_grain}
\end{figure*}

\subsection{Slowly-varying transverse modulations}

We are now able to provide a modulational description of transverse wavenumber dynamics in \eqref{e:sh-c}. We shall once again focus on parallel striped fronts, fix a $c>0$ where parallel striped fronts $u_\rf(\xi,\omega t;k_x)$ exist, and are diffusively stable. Fixing $\xi=\xi_0<0$ just behind the quenching line, we consider a modulation of the traveling front solution through the ansatz, $u_\rf(\xi,\Phi(Y,T)+\omega t;\delta\Phi_Y(Y,T)) $. Inserting and expanding in $\delta$, we once again obtain a viscous Burgers' modulation equation \eqref{e:vb} for the transverse wavenumber $\Psi$ with the coefficients $\lambda_\rlin''(0)$ and $\omega_\rf''(0)$ found above.  In all numerical simulations of \eqref{e:vb} discussed below, we use the numerical approximations of these coefficients as described above.

 Using this modulation equation, we obtain accurate predictions for the transverse wavenumber evolution. In Figure \ref{f:sh_mod_num} we consider a localized phase perturbation of a weakly oblique stripe of the same form as in \eqref{e:loc_in}.  To measure the transverse wavenumber we use the Hilbert transform 
$$
\mathcal{H}[u](y) := \mathcal{F}^{-1}\left[ -\ri\, \mathrm{sign}(\ell) \mathcal{F}[u](\ell)\right](y),
$$
 where $\mathcal{F}[u](\ell)$ denotes the Fourier transform in $y$ and $\ell$ the Fourier wavenumber variable, to construct a complex signal $v(y) = u(y)+ \ri \mathcal{H}[u](y)$ which, for oscillatory functions $u(y)\approx \cos(\phi(y))$, can be differentiated to obtain a local wavenumber $\psi(y,t) =\mathrm{Im}\, \partial_y v/v $. As the Hilbert transform induces spurious oscillations in the measured wavenumber, we use the iterative transform approach of (Ref. \onlinecite{gengel}) to reduce, though not eliminate, the occurrence and magnitude of such oscillations.   Once again using a scaled version of the initial wavenumber,  $\delta^{-1}\psi(Y/\delta,0)$, as the initial data for the Burgers' equation \eqref{e:vb}, we find good agreement between the predicted and measured wavenumbers, with errors behaving similarly as those in CGL.




In Figure \ref{f:sh_mod_grain} we consider a pair of symmetric grain boundary defects which connect weakly oblique stripes of the opposite wavenumber $k_y = \pm\delta$, with one grain boundary convex to the quench line and the other concave. The convex grain, located in the middle of the computational domain, has left wavenumber $k_{y,-} = -\delta$ and right wavenumber $k_{y,+} = +\delta$. Since, $\omega_\rf''(0)<0$, this defect behaves as a shock-like sink in the transverse direction, with speed $c_* = \omega''(0)(k_{y,-}+k_{y,-})/2 = 0$ and wavenumber interface which remains sharply localized. The concave grain, located near the boundary of the periodic computational domain, has ``left" wavenumber $k_{y,-} = \delta$ and ``right" wavenumber $k_{y,+} = -\delta$. Thus the corresponding solution to the Burgers' equation has outward pointing characteristics and thus behaves like a rarefaction wave. 

 We remark here that wavenumber measurements using the Hilbert transform (which in MATLAB uses the discrete Fourier transform) induce Gibbs-type oscillations due to the sharp viscous shock profile in the wavenumber, leading to larger errors in the part of the domain where the shock resides.  We remark that we also studied an asymmetric grain boundary in \eqref{e:sh-c} with moving shock and rarefaction wave as studied in CGL and depicted in Figure \ref{f:sink-source} above. While not depicted, we obtained accurate shock speed predictions from the viscous Burgers modulational approximation.


\section{Conclusion and discussion }\label{s:conc}
To summarize, we derived a one-dimensional viscous Burgers' modulation equation to describe small-amplitude transverse wavenumber dynamics of asymptotically-stable striped patterns in the wake of a rigidly propagating directional quench. Of most interest, we found that the nonlinear flux coefficient, $\omega_\rf''(0)$, is determined through the wavenumber selection properties of the directional quench. Somewhat less surprisingly, the viscosity coefficient, $\lambda_\rlin''(0)$, in the modulation equation is determined by the transverse diffusivity of pure stripes.  This modulation equation accurately predicted finite-time dynamics of small amplitude wavenumber defects just behind the quenching line, including source-sink pairs and localized phase-slip defects, in both the complex Ginzburg-Landau and Swift-Hohenberg equations. While we only considered examples of coherent defects, we expect the modulation equation to give accurate predictions for arbitrary wavenumber modulations which are smooth and small-amplitude. Furthermore, we expect this modulational analysis to accurately predict the dynamics of directionally quenched stripes in general dissipative systems where the underlying asymptotic pattern is diffusively stable and the quench selects wavenumbers. That is, for a fixed quenched speed $c$ and transverse wavenumber $k_y$, the horizontal wavenumber $k_x$ is locally unique.

There are several areas of further study to extend from this work. The first and most natural next step would be to consider the far-field dynamics by deriving a two-dimensional modulational equation for the striped phase or wavenumber in the half-plane to the left of the quenching line at $x = ct$, with vertical boundary condition along the $y$-axis. One would seek to derive a Hamilton-Jacobi equation (Ref. \onlinecite{howard77}) or Cross-Newell equation (Refs. \onlinecite{passot1994towards,ercolani2000geometry}) for the phase dynamics, possibly through an intermediate Newell-Whitehead-Segel amplitude equation (Ref. \onlinecite{malomed1990domain}).  Of most interest would be to determine a suitable boundary condition for the phase to represent the quenching line.  In the limit of slow quenching speed, the work (Ref. \onlinecite{chen2021strain}) used a linear phase-diffusion equation $\varphi_t = \Delta \varphi+c\varphi_x$, with nonlinear boundary condition $\varphi_x = g(\varphi), \quad (x,y)\in \{0\}\times \R$ on the vertical axis, to describe the dynamics of the phase $\varphi$. Here the nonlinear boundary condition is determined by an object known as the \emph{strain-displacement} relation of stationary quenched stripes, a curve which parameterizes the set of possible wavenumbers selected by the quench in terms of the asymptotic phase. We do not expect such a model to be valid for intermediate or fast growth. To our knowledge, an appropriate boundary condition has not been derived for these regimes. Such a phase description would allow one to understand the precise far-field behavior of the defects observed above. For example, they would allow one to describe how the Swift-Hohenberg grain-boundary formed in Figure \ref{f:sh_mod_grain} relaxes or evolves as $x\rightarrow-\infty$. After obtaining such a two-dimensional approximation, it would also be of interest to obtain rigorous approximation results, as given in (Ref. \onlinecite{dmwt}), between the modulation equation and the full system. 

Other avenues of subsequent study include extending this analysis to three spatial dimensions $(x,y,z)\in\R^3$ with a planar quench propagating in the $x$-direction and a modulation equation for the two-dimensional transverse dynamics in $(y,z)$. Finally, it would be of interest to derive transverse modulation equations for non-directional quenching where the interface bounding the pattern-forming regime is not a hyperplane, but an evolving curve or sub-manifold.


\begin{acknowledgments}
The authors were partially supported by the National Science Foundation through grant NSF-DMS-2006887. 
\end{acknowledgments}

\section*{Data Availability Statement}
The data and code that support the findings of this study are available from the corresponding author upon reasonable request.

\section*{Author Contributions}
\textbf{Sierra Dunn}: writing – review and editing (equal); 
 numerical computations (equal); formal analysis (supporting).
\textbf{Ryan Goh}: Conceptualization (lead); writing - original draft (lead); writing – review and editing (equal); formal analysis (lead); numerical computations (equal).
\textbf{Benjamin Krewson}: writing – review and editing (equal); numerical computations (equal); formal analysis (supporting).

\appendix
\section{Derivation of transverse modulation equation}\label{a:1}

Below we provide the derivation of the transverse modulation equation \eqref{e:vb} for the quenched CGL equation \eqref{e:cgl-c}. An analogous approach gives the same modulation equation for the quenched SH equation \eqref{e:sh-c}.

We seek to modulate the traveling wave solutions $A_*(\xi,y,t;k_y):= \re^{\ri (k_y y - \omega t)} A_\mathrm{f}(x;k_y)$ of \eqref{e:cgl-c} described above, where $A_\mathrm{f}$ is a solution of the traveling wave equation \eqref{e:tw}. We write the latter equation in the condensed form
\begin{align}
0 &= L(k_y) A + N(A,\bar A),\qquad \label{e:CLN0}\\
& L(k_y) = (1+\ri\alpha) (\partial_\xi^2-k_y^2) + c\partial_\xi + \chi-\ri\omega_\mathrm{f}(k_y),\notag\\
&N(A,\bar A) =-(1+\ri\gamma) A^2\bar A =-(1+\ri\gamma) A|A|^2,\notag
\end{align}
where we recall that $\omega_\mathrm{f}(k_y)$ is the selected frequency of the quenched front determined by $k_y$. 

Note, to obtain a smooth equation, we consider $A$ and $\bar A$ independently,  and hence must also consider the complex conjugate of \eqref{e:CLN0}. Taken together, denoting $U = (A,\bar A)^T$, we thus consider
\begin{eqnarray}\label{e:CLN1}
&\left(\begin{array}{c}0 \\0\end{array}\right) = \mathcal{F}(U;k_y) := 
\mathcal{L}(k_y) U + \mathcal{N}(U),\qquad\\
&\mathcal{L}(k_y):= \left(\begin{array}{cc}L(k_y) & 0 \\0 & \bar L(k_y)\end{array}\right),\,\, \mathcal{N} = \left(\begin{array}{c} N(U) \\ \bar N(U)\end{array}\right).
\end{eqnarray}

\subsection{Transverse wavenumber dependence and stability of CGL front}\label{ss:aky}
  Before performing the modulation expansion, we consider some properties of \eqref{e:CLN1}, its dependence on $k_y$, and its linearization.  To begin, we evaluate the equation on the front, $0 = \mathcal{F}(U_\mathrm{f}(\cdot;k_y);k_y)$ and differentiate in $k_y$ to first and second order, obtaining
\begin{align}
0 &= \left[\mathcal{L}(k_y) + D\mathcal{N}(U_\mathrm{f}) \right] \partial_{k_y} U_\mathrm{f}(k_y) + \partial_{k_y}\mathcal{L}(k_y)U_\mathrm{f}(k_y),\label{e:dky}\\
0 &=  \left[\mathcal{L}(k_y) + D\mathcal{N}(U_\mathrm{f}) \right]  \partial_{k_y}^2 U_\mathrm{f} (k_y)+2 \partial_{k_y}\mathcal{L}(k_y) \partial_{k_y} U_\mathrm{f}(k_y)\notag \\
&\quad+D^2\mathcal{N}(U_\mathrm{f})\left[\partial_{k_y} U_\mathrm{f}(k_y), \partial_{k_y} U_\mathrm{f}(k_y) \right].\label{e:dky2}
\end{align}
with Jacobian
$$
D\mathcal{N}\left(\begin{array}{c} A\\ \bar A\end{array}\right) = 
-\left(\begin{array}{cc}2(1+\ri\gamma) A \bar A  & (1+\ri\gamma) A^2 \\ (1-\ri\gamma) \bar A^2 & 2(1-\ri\gamma) A\bar A\end{array}\right), \quad 
$$
and Hessian quadratic form
\begin{eqnarray*}
D^2\mathcal{N}(U_\mathrm{f})\left[ U,V\right]= U^T \left(\begin{array}{cc}-2(1+\ri\gamma) \bar A_\mathrm{f} & -2(1+\ri\gamma) A_\mathrm{f} \\ -2(1-\ri\gamma) A_\mathrm{f} & -2(1-\ri\gamma) \bar A_\mathrm{f}\end{array}\right) V.
\end{eqnarray*}
Evaluating at $k_y = 0$ and recalling from \eqref{e:cgl_disp} above that $\partial_{k_y}\omega(0) = 0$, we have that $\partial_{k_y}\mathcal{L}(0) = 0$ and 
\begin{eqnarray*}
&\partial_{k_y}^2\mathcal{L}(0) = 
 \left(\begin{array}{cc}-2(1+\ri\alpha) & 0 \\0 & -2(1-\ri\alpha) \end{array}\right) + 
\partial_{k_y}^2\omega_\mathrm{f}(0)\left(\begin{array}{cc}-\ri & 0 \\0 &  \ri\end{array}\right).
\end{eqnarray*}
Next, the gauge action $(A,\bar A)\mapsto (\re^{\ri\theta} A, \re^{-\ri\theta} \bar A)$, induces a $0$-eigenvalue with eigenfunction $U_0:= (\ri A_\mathrm{f}, -\ri A_\mathrm{f})^T$ of the linearization $$\mathbb{L}:= \mathcal{L}(0) + D\mathcal{N}(U_\mathrm{f}(0))$$ of $\mathcal{F}$ at $(U_\mathrm{f}(\cdot; \, 0),0)$, defined in an exponentially weighted function space with growing weights at $\xi = \pm\infty$.
Due to the lack of $\xi$-translational invariance caused by the inhomogeneous quenching term,  we find $\ker \mathbb{L} = \mathrm{span}\{ U_0\}$.  One also can define the formal adjoint of $\mathbb{L}$ as 
$$
\mathbb{L}^\mathrm{ad}:= \left(\begin{array}{cc}L(0)^\mathrm{ad} & 0 \\0 & L(0)^\mathrm{ad}\end{array}\right) + D\mathcal{N}(U_\mathrm{f})^*,
$$
where $L(0)^\mathrm{ad} = (1-\ri\alpha)\partial_\xi^2 -c\partial_\xi + \chi - \ri\omega_\rf$ and $*$ denotes the complex-conjugate transpose of a matrix. We also let $U^\mathrm{ad}$ denote the element spanning $\ker \mathbb{L}^*$ which satisfies $\la U^\mathrm{ad},U_0\ra_{L^2(\R)^2} = 1.$

Evaluating \eqref{e:dky} at $k_y = 0$ we find that $\partial_{k_y} U_\mathrm{f}$, if it is non-trivial, must lie in $\ker \mathbb{L}$. Evaluating \eqref{e:dky2} at $k_y = 0$ and moving the term involving $\partial_{k_y}^2\mathcal{L}(0)$ over to one side, we obtain
\begin{eqnarray}\label{e:lky2}
&\partial_{k_y}^2\mathcal{L}(0) U_\mathrm{f} = -\Big[\left( \mathcal{L}(0) + D\mathcal{N}(U_\mathrm{f})\right)\partial_{k_y}^2U_\mathrm{f}\notag\\
& + D^2\mathcal{N}(U_\mathrm{f})[\partial_{k_y}U_\mathrm{f}, \partial_{k_y} U_\mathrm{f}]  \Big].
\end{eqnarray}
 Taking the $L^2(\mathbb{R})$ inner product with $U^\mathrm{ad}$, we then obtain
\begin{align}\label{e:dkyw}
& - \partial_{k_y}^2 \omega_\rf(0) =  \la U^\mathrm{ad}, - \partial_{k_y}^2 \omega(0) U_0\ra\notag\\
 & = 2\la U^\mathrm{ad}, \left(\begin{array}{c} (1+\ri\alpha) A_\mathrm{f} \\ (1-\ri\alpha)\bar A_\mathrm{f} \end{array}\right)\ra - \la U^\mathrm{ad}, D^2\mathcal{N}(U_\mathrm{f}) [\partial_{k_y}U_\mathrm{f}, \partial_{k_y} U_\mathrm{f}]  \ra. 
 \end{align}
In principle, one could numerically approximate the inner products on the right hand side. As described in the main body of the text, we numerically continue front-wavenumber pairs $(A_\rf,k_{x,\rf})$ in $k_y$. This allows us to estimate derivatives of the curve $k_{x,\rf}$ and thus compute $\omega_\rf''(0)$. 

To obtain the viscosity parameter of the modulation equation, we also need to consider the linearized dynamics of transverse perturbations of the parallel-striped front near the interface.
To begin, we consider transversely modulated perturbations of the parallel striped front $A = \re^{\ri\omega t}\left(A_\rf(\xi;0) + a(\xi,t)\re^{\nu y} \right),\,\, \nu\in\ri\R$ in equation \eqref{e:cgl-c}, and obtain at the linear level in $a$ and $\bar a$, after including the complex conjugate equation, the system
\begin{eqnarray}
&V_t = \widetilde{\mathbb{L}}(\nu) V, \quad  \\
&V = (a,\bar a)^T,\quad \widetilde{\mathbb{L}}(\nu) =  \left(\begin{array}{cc}\tilde L(\nu) & 0 \\0 & \overline{\tilde L}(\nu)\end{array}\right) + D\mathcal{N}(U_\rf),\notag
\end{eqnarray}
where $\tilde L(\nu) = (1+\ri\alpha)(\partial_\xi^2 +\nu^2) + c \partial_\xi + \chi - \ri \omega_\rf(0)$ (compare to the operator $L(k_y)$ in \eqref{e:CLN0}). We then consider transversely modulated eigenvalues 
$$
\widetilde{\mathbb{L}}(\nu) V(\nu) = \lambda_\rlin(\nu) V(\nu),
$$
where $ V(0) = U_0$ gives the gauge-action eigenfunction with eigenvalue $\lambda_\rlin(0) = 0$ discussed above. In a similar manner to the $k_y$ dependence of the front in Section \ref{ss:aky}, we twice-differentiate the eigenvalue equation with respect to $\nu$, evaluate at $\nu= 0$, use the fact that $\partial_\nu \tilde{\mathbb{L}}(0) = 0$, and take the inner product with $U_\mathrm{ad}$ to obtain
\beq\label{e:lam2lin}
\lambda''_\rlin(0) = 2 \la \left(\begin{array}{c}\ri (1+\ri\alpha) A_0\\-\ri(1-\ri\alpha) A_0\end{array}\right) , U_\mathrm{ad} \ra.
\eeq
We approximate $\lambda_\rlin''(0)$ by considering transverse perturbations of a pure parallel-stripe. Following (Ref. \onlinecite{CGL_Lambda_Stability}), we perturb stripes with the ansatz $A(x,y,t) = \re^{\ri\omega t} \left[ r_\rf \re^{\ri k_\rf x} + a_+ \re^{\lambda t + \nu y} + a_- \re^{\bar \lambda t - \nu y} \right]$ in \eqref{e:cgl-c}, collecting $\mathcal{O}(a_\pm)$ terms, solving for $\lambda$, and expanding in $\nu\sim0$ to obtain
\beq\label{e:lamlin}
\lambda_\rlin(\nu) = (1+\alpha\gamma)\nu^2 -\frac{\alpha^2}{2}(1+\gamma^2)\nu^4 + \mc{O}(\nu^6), \quad \nu\in\ri\R.
\eeq
Importantly, our assumption that the selected asymptotic waves are Benjamin-Feir stable, $1+\alpha\gamma>0$, gives $\lambda_\rlin''(0) >0$.


\subsection{Modulational ansatz and expansion}
As described above, we consider the modulational ansatz
$$
A(x,y,t) = A_\mathrm{f}(x; \delta \Phi_Y) \re^{\ri(\Phi - \omega t)}, 
$$
with $\Phi = \Phi(Y,T)$ a long-wavenlength phase modulation function of the variables $Y = \delta y$ and $T = \delta^2 t$ for $0<\delta\ll1$. We then expand 
\begin{align}
A_\mathrm{f}(x;\delta\Phi_Y) &= A_\mathrm{f}(x;0) + \delta \Phi_Y \partial_{k_y} A_\mathrm{f}(x;0)\notag\\
&\qquad\qquad + \frac{\delta^2}{2} \Phi_Y^2 \partial_{k_y}^2 A_\rf(x;0) + \mathcal{O}(\delta^3).\notag
\end{align}
Note, to ease notational burden, we let $A_0 = A_\mathrm{f}$ and $A_1 = \partial_{k_y} A_\mathrm{f}$. Before inserting this expansion into \eqref{e:cgl-c}, we calculate several derivatives of the expanded ansatz $A_\mathrm{f}(x; \delta \Phi_Y) \re^{\ri(\Phi - \omega t)}$:
\begin{align}
\partial_t A(x,y,t) &= -\ri\omega A_0 - \delta \ri\omega \Phi_Y A_1 \notag\\
&+ \delta^2\left( \ri \Phi_T A_0 - \ri \frac{\omega}{2} \Phi_Y^2 A_2 \right) + \mathcal{O}(\delta^3),\notag\\
\partial_y^2 A(x,y,t) &= \delta^2\left( \ri \Phi_{YY} A_0 -\Phi_Y^2 A_0\right) + \mc{O}(\delta^3).\notag
\end{align}
The cubic nonlinearity expands as
\begin{eqnarray}
&A^2 \bar A = A_0^2 \bar A_0 + \delta\left( 2A_0\bar A_0 A_1 \Phi_Y + A_0^2\bar A_1 \Phi_Y\right)\notag\\
&+\delta^2\Big( A_0\bar A_0 A_2 \Phi_Y^2 + \frac{A_0^2}{2} \bar A_2 \Phi_Y^2 \notag\\
&\qquad + \bar A_0 A_1^2 \Phi_Y^2 + 2 A_0 \Phi_Y^2 A_1 \bar A_1\Big)+\mathcal{O}(\delta^3),
\end{eqnarray}
while the expansion for $A\bar A^2$ is obtained by taking complex conjugates. Inserting the expanded ansatz into the full PDE \eqref{e:cgl-c} and separating out orders of $\delta$, we obtain at $\mc{O}(1)$ the traveling wave equation \eqref{e:CLN1} for $k_y = 0$. At $\mc{O}(\delta)$ we obtain 
$$
0 = \mathbb{L} \left(\begin{array}{c}\Phi_Y A_1 \\ \Phi_Y \bar A_1\end{array}\right) = \Phi_Y  \mathbb{L} \left(\begin{array}{c} A_1 \\  \bar A_1\end{array}\right),
$$
as $\Phi_Y$ is independent of $x$. Note this is consistent with \eqref{e:dky} above. Finally, at $\mc{O}(\delta^2)$ we obtain 
\begin{widetext}
\begin{align}
\ri \Omega_T A_0 - \frac{\omega}{2} \Phi_Y^2 A_2  &= (1+\ri\alpha) \left( \frac{\Phi_Y^2}{2} \partial_x^2 A_2  + (\ri\Phi_{YY} - \Phi_Y^2)A_0\right) + (c\partial_x + \chi)\frac{\Phi_y^2}{2} A_2\notag\\
&\qquad\qquad\qquad - (1+\ri\gamma) \left( A_0 \bar A_0 A_2 \Phi_Y^2 + A_0^2 \bar A_2 \frac{\Phi_Y^2}{2} + \frac{\Phi_Y^2}{2} (2 \bar A_0 A_1^2 + 4 A_0 A_1 \bar A_1) \right)
\end{align}
\end{widetext}
and its complex conjugate equation. Rearranging, and using \eqref{e:lky2} we then find
\begin{align}
&\left(\begin{array}{c}\ri A_0\left(\Phi_T - (1+\ri\alpha)(\Phi_{YY} + \ri \Phi_Y^2) \right) \\ -\ri A_0\left(\Phi_T - (1-\ri\alpha)(\Phi_{YY} - \ri \Phi_Y^2) \right)\end{array}\right) \notag\\
&=\frac{\Phi_Y^2}{2}\mathbb{L} \left(\begin{array}{c}A_2\\\bar A_2\end{array}\right)+
\frac{\Phi_Y^2}{2} (A_1,\bar A_1)D^2\mathcal{N}(A_0,\bar A_0) \left(\begin{array}{c}A_1\\ \bar A_1\end{array}\right)\notag\\
&= -\frac{\Phi_Y^2}{2} \partial_{k_y}^2\mathcal{L}(0) \left(\begin{array}{c}A_0 \\\bar A_0\end{array}\right)\notag\\
&= \frac{\Phi_Y^2}{2}   \left(\begin{array}{c} 2(1+\ri\alpha )A_0 \\ -2(1-\ri\alpha)\bar A_0\end{array}\right) +  \frac{\Phi_Y^2}{2}\partial_{k_y}^2 \omega_\rf(0) \left(\begin{array}{c}\ri A_0\\-\ri A_0\end{array}\right). 
\end{align}
Using \eqref{e:lam2lin}, this can then be simplified to obtain
\begin{align}
&\left(\begin{array}{c}\ri A_0\left(\Phi_T - (1+\ri\alpha)\Phi_{YY} \right) \\ -\ri A_0\left(\Phi_T - (1-\ri\alpha)\Phi_{YY}) \right)\end{array}\right) =\frac{\Phi_Y^2}{2} \partial_{k_y}^2 \omega_\rf(0) \left(\begin{array}{c}\ri A_0\\-\ri A_0\end{array}\right).
\end{align}

Taking the inner product of this last equation with $U^\mathrm{ad}$ and using the fact that $U_0 = (\ri A_0, -\ri A_0)^T$ we then obtain
\begin{align}\label{e:phimod}
\Phi_T = \la \left(\begin{array}{c}\ri (1+\ri\alpha) A_0\\-\ri(1-\ri\alpha) A_0\end{array}\right) , U_\mathrm{ad} \ra \Phi_{YY} + \frac{\partial_{k_y}^2 \omega_\rf(0)}{2} \Phi_Y^2.
\end{align}
Combining this with the computation in \eqref{e:lam2lin}, we obtain from \eqref{e:phimod} the desired leading order phase modulation equation,
\beq
\Phi_T = \frac{\lambda''_\rlin(0)}{2} \Phi_{YY} + \frac{\omega_\rf''(0)}{2}\Phi_Y^2,
\eeq
which can be readily differentiated in $Y$ to obtain the equation for the wavenumber modulation given in \eqref{e:vb} above.

\bibliography{transv_mod}

\end{document}